\documentclass[10pt,twocolumn,twoside,openbib]{article}
\usepackage[footnotesize]{caption} \newcommand{\icar}[1]{} \newcommand{\prep}[1]{#1} \prep{\setlength{\textwidth}{ 18cm}} \prep{\setlength{\textheight}{  22.cm}} \prep{\setlength{\oddsidemargin}{  -1cm}} \prep{\setlength{\evensidemargin}{ -1cm}} \prep{\setlength{\topmargin}{  -2cm}}

\usepackage{amssymb,amsmath,mathrsfs}
\usepackage{multirow}
\usepackage{graphicx}
\usepackage{float}
\usepackage{color}
\usepackage{ulem}
\usepackage{natbib}
\def \eps {\varepsilon}

\def\etal{{et al.}}

\newcommand{\blanc}[1]{}

\newcommand{\ZZ}{{\mathbb Z}}
\newcommand{\NN}{{\mathbb N}}
\newcommand{\cA}{{\cal A}}
\newcommand{\cB}{{\cal B}}
\def\cC{{\cal C}}
\def\cS{{\cal S}}
\def\alp{{\alpha}}
\def\lam{{\lambda}}

\def\eps{{\varepsilon}}
\def\ab{{\bar a}}
\def\nb{{\bar n}}
\def\nc{{\Check n}}
\def\sigb{{\bar \sigma}}
\def\eb{\bar e}

\def\be{\begin{equation}}
\def\ee{\end{equation}}
\def\Dron#1#2{\frac{\partial#1}{\partial#2}}

\def\Der#1#2{\frac{d#1}{d#2}}

\def\etal{{\it et~al. }}

\def\figureXa
{\begin{figure}[ [!ht] 
   \centering
   \includegraphics[width=8cm]{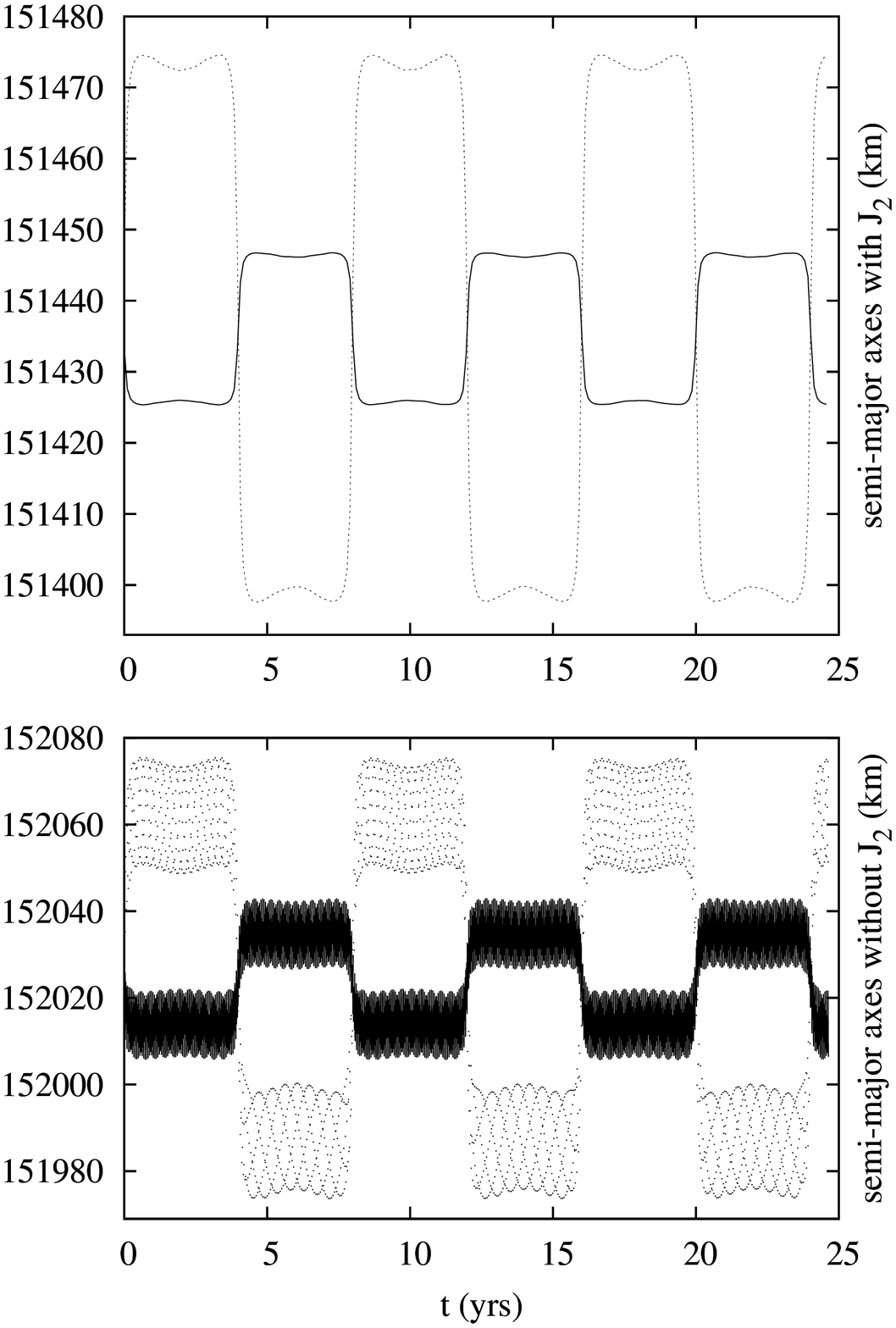} 
\caption{Variation of the semi-major axis of Janus and Epimetheus: the upper plot shows the temporal variation of the semi-major axes of the two satellites (solid line for Janus and dotted line for Epimetheus), when $J_2$ is included in the third Kepler's law. The semi-major axes oscillate around the mean value $\ab = 151436.9$ km.  When $J_2$ is not considered (bottom plot), large short-period oscillations are superimposed to the above-mentioned $8$-year signal. The signals have a period of $2\pi/\nu \approx 8\, {\rm years}$.  In the latter case, the mean value of $a$ is equal to $152024.4$ km, which is  about $600$ km greater than when $J_2$ is taken into account. }
   \label{fig:a}
\end{figure}
}

\def\figureXrelat
{
\begin{figure}[!ht] 
   \centering
   \includegraphics[width=8.5cm]{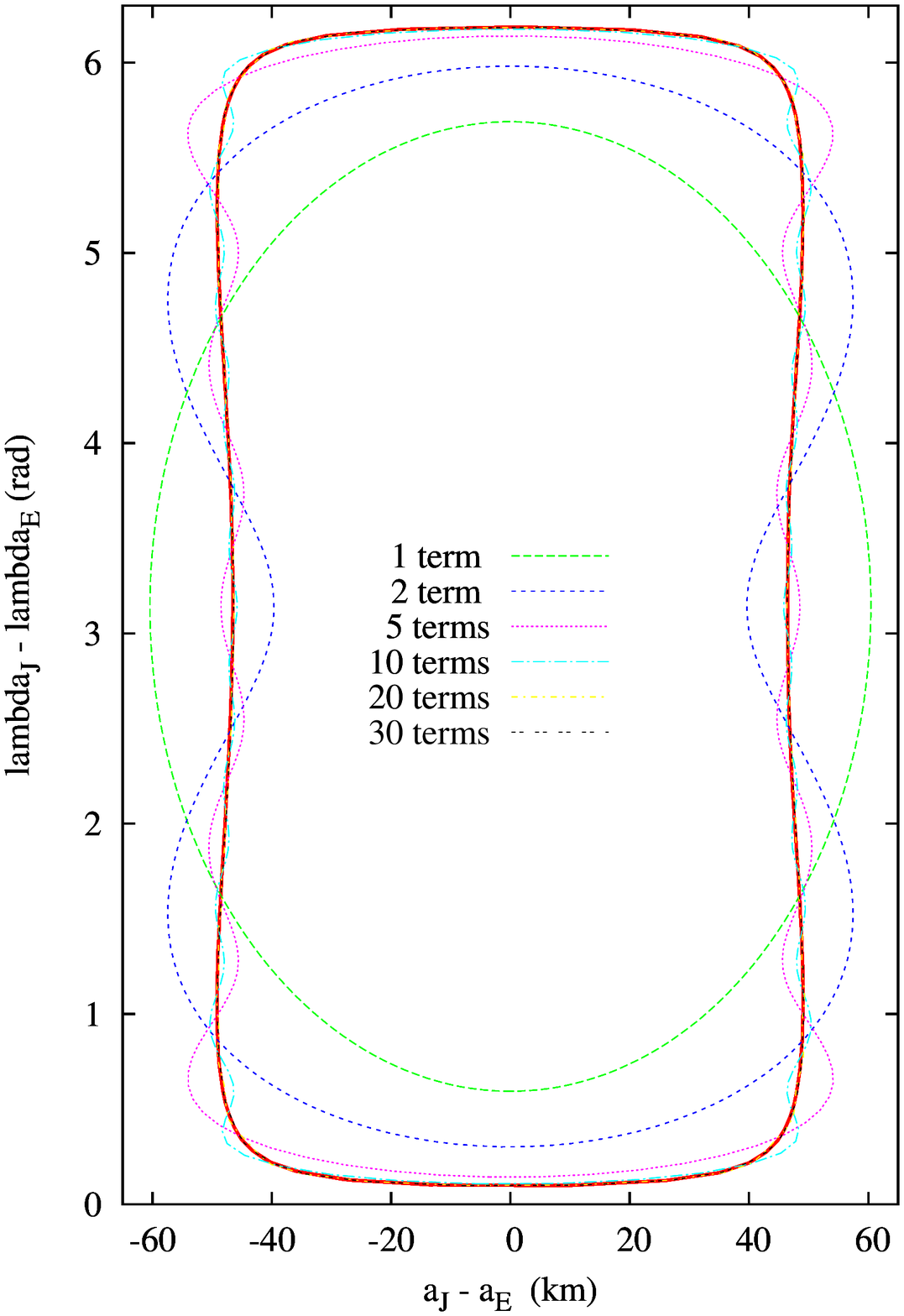} 
  \caption{Relative orbit of the satellites:  the X-axis represents the relative semi-major axis $a_r = a_J-a_E$ in km, while the Y-axis corresponds to the relative mean longitude $\lam_r = \lam_J-\lam_E$ in radians. The solid red curve plots the orbit deduced from the numerical simulations, while the dotted curves stands for several approximation given by formulas~(\ref{eq:arelat}) and~(\ref{eq:larelat}). See text for more details. }
   \label{fig:QP_relat}
\end{figure}
}

\def\figureXecc
{
\begin{figure}[!htbp]
\begin{center}
   \includegraphics[width=6.3cm, angle = -90]{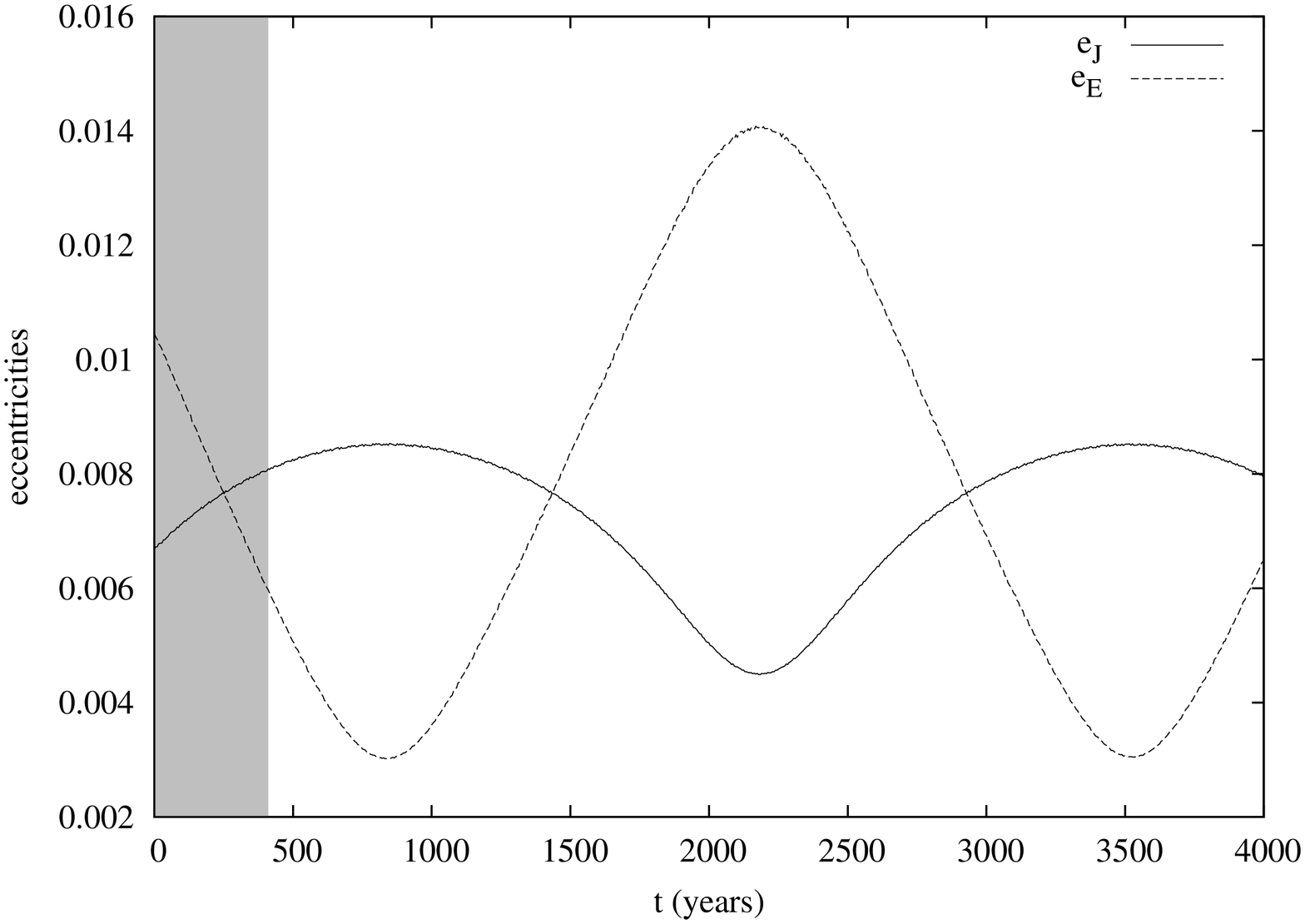}
\caption{Secular variations of the eccentricities of the two satellites (solid curve for Janus, dashed curve for Epimetheus) due to their mutual gravitational interactions. The grey rectangle represents the time interval used for the numerical simulation of the rotation.}
\label{fig:ecc}
\end{center}
\end{figure}
}

\def\figureXerr
{
\begin{figure}[!ht]
\begin{center}
   \includegraphics[width=8cm]{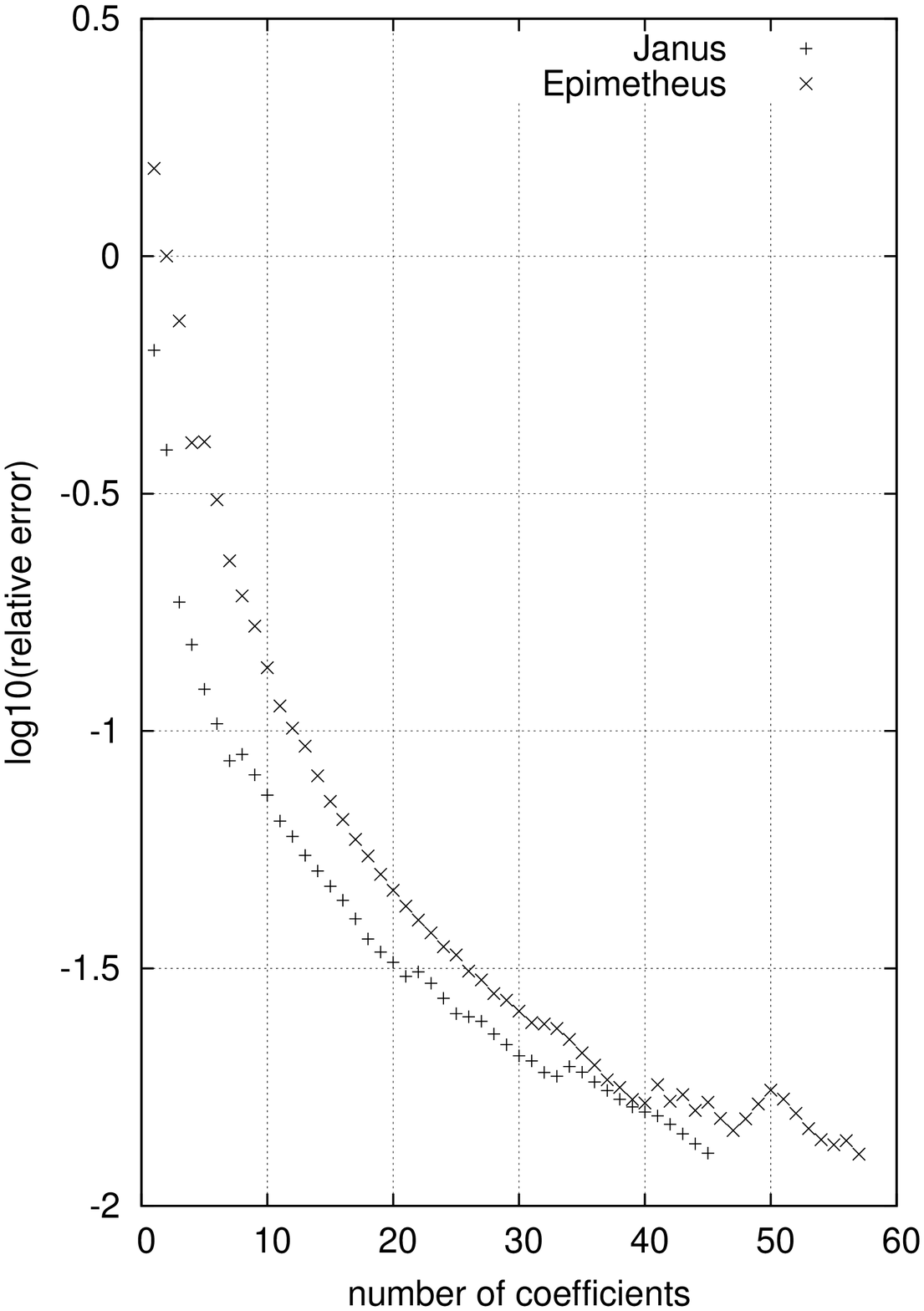}
\caption{ Relative error $\epsilon_p$ of the Fourier approximations of $\tilde\gamma$ against the number of terms included in the decomposition. The $"+"$ correspond to Janus and the $"\times"$ to Epimetheus.}
\label{fig:err_coef}
\end{center}
\end{figure}
}

\def\figureXerrE
{
\begin{figure}[!ht]
\begin{center}
   \includegraphics[width=8cm]{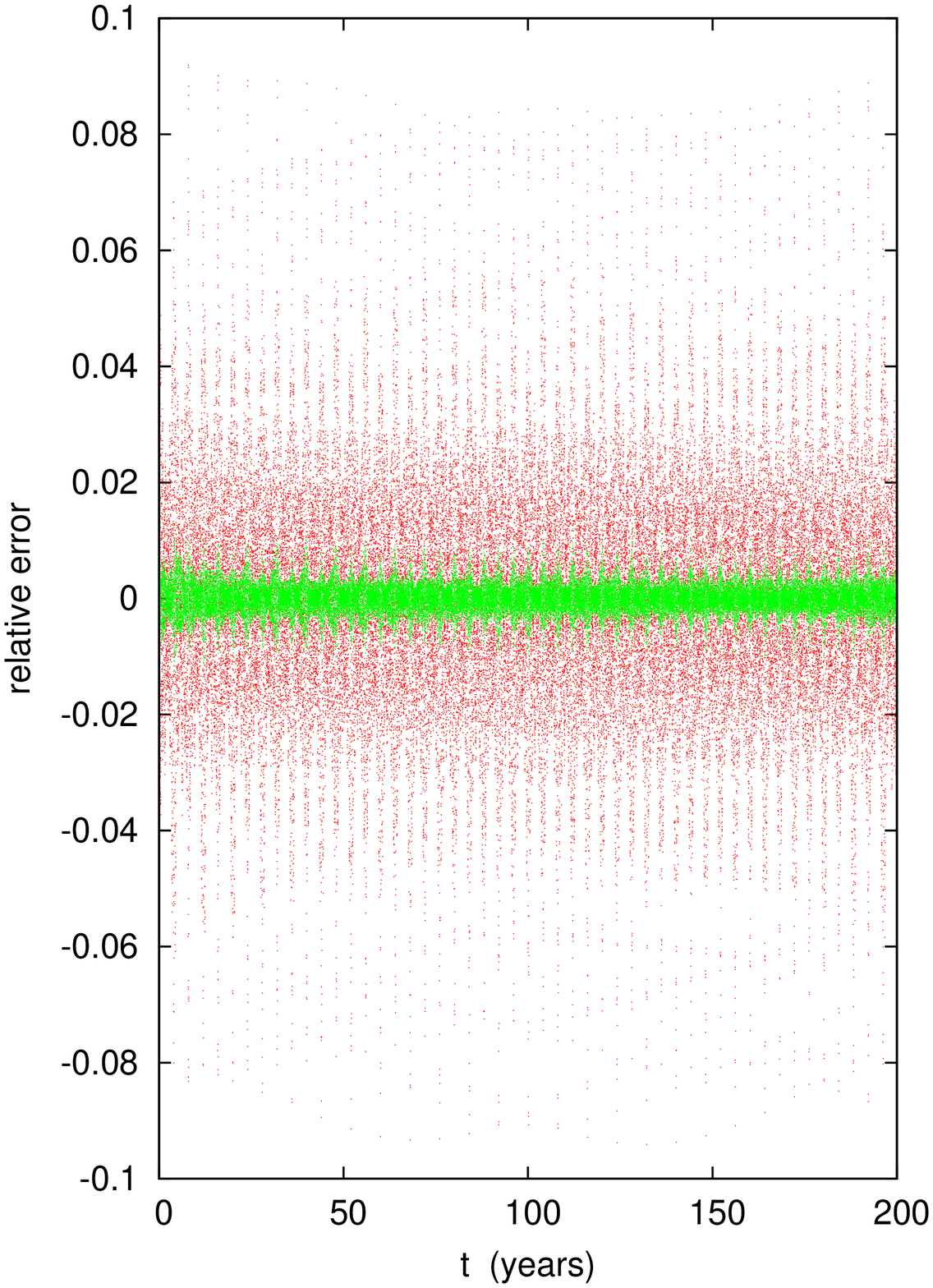}
\caption{Relative error of the Fourier approximation of $\tilde\gamma$ in the interval $[\!\,0\!:\!200\!\,]\, \text{years} = [JD\, 2433278.5:JD\, 2433278.5]$ for Epimetheus. The instantaneous relative error $\left( \tilde\gamma_s(t) - \tilde\gamma_s^p(t)\right) / {\rm Max}\vert  \tilde\gamma_s(t)\vert$ is plotted in red for $p=13$  and  in green for $p=57$.}
\label{fig:err_Epim}
\end{center}
\end{figure}
}

\def\figureXtriax
{
\begin{figure}[!ht] 
   \centering
   \includegraphics[width=6.cm, angle=-90]{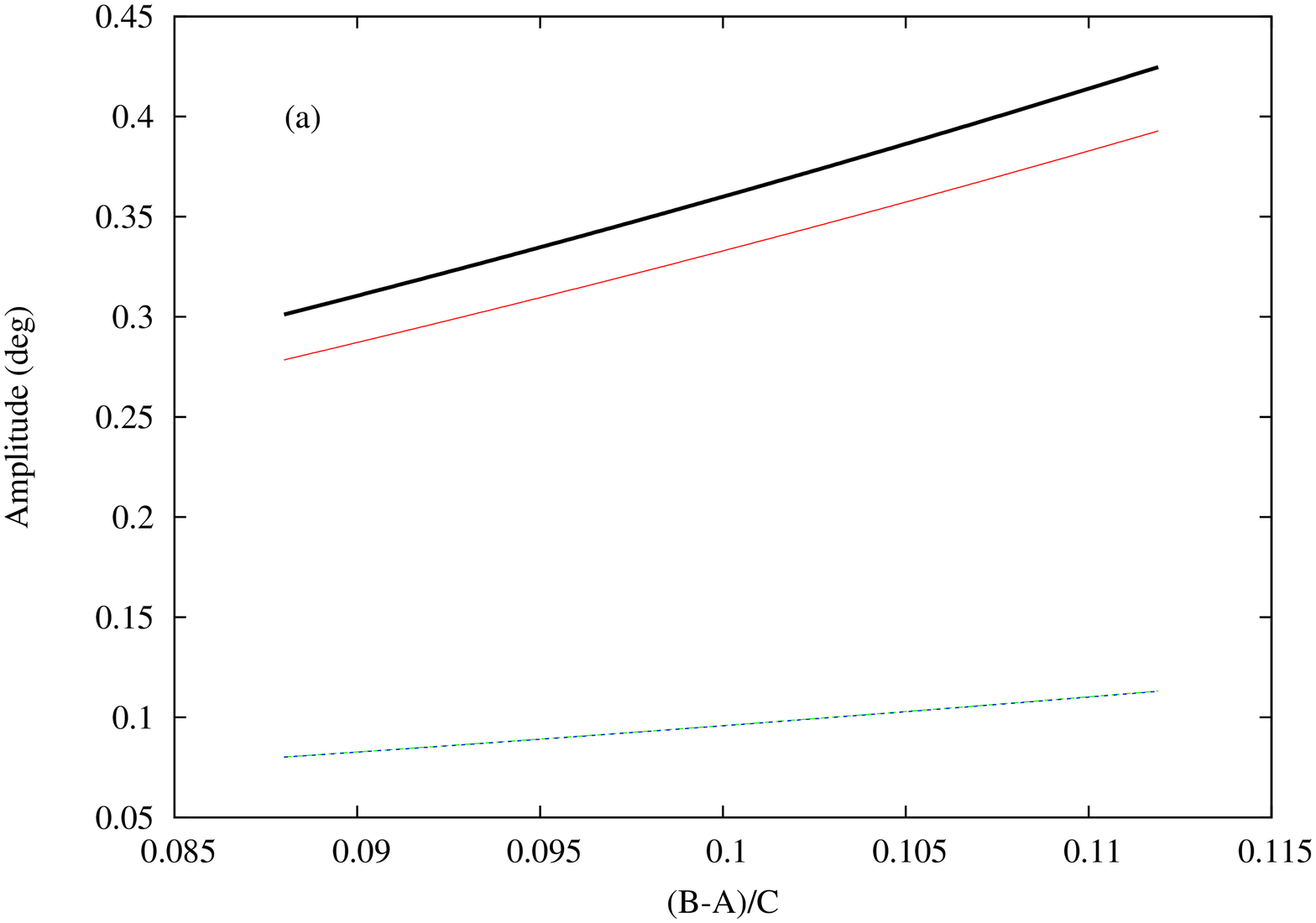} 
   \includegraphics[width=6.cm, angle=-90]{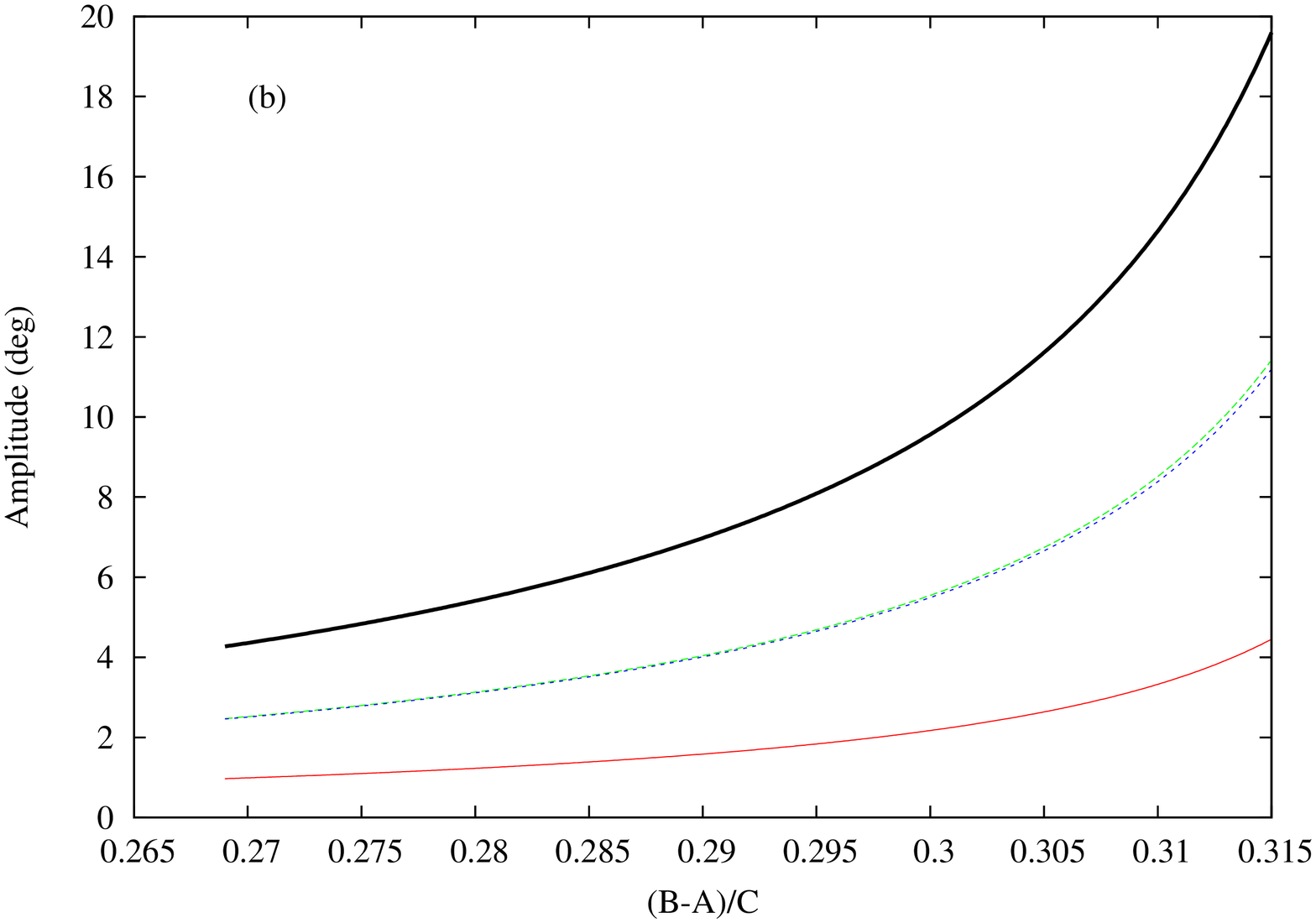} 
   \caption{Influence of the triaxiality on the amplitude of the mode $\nb-g_k$ (red curve) where $k$ is equal to $J$ for Janus (a) and $E$ for Epimetheus (b), $\nb-g_k + \nu$ (green curve), and $\nb-g_k - \nu$ (blue) from Eq.~(31). The black bold curves correspond to the amplitude deduced from the adiabatic invariant model (Eq.~\ref{eq:sol0}).}
   \label{fig:triax}
\end{figure}
}

\def\tableXfrefond
{\begin{table}
\caption{Fundamental frequencies characteristic of the Saturn-Janus-Epimetheus system. These frequencies are derived from a $400$ years  long numerical integration of the three-body problem including Saturn's oblateness.}
\begin{center}
\begin{tabular}{crl }
       &  Freq. (rad/day)   & Per.   \\
\noalign{\smallskip}
\hline
\hline
  $\nb$ &  9.045924658  &   0.69459 (days)  \\ 
 $\nu$ &  -0.002147139  &   8.01179 (yrs)  \\ 
 $g_J$ &  0.034948139  &   0.49223	(yrs)  \\ 
 $g_E$ & 0.034952723   &   0.49216 (yrs)\\ 
 $s_J$ & -0.034811959   &   0.49415 (yrs) \\ 
 $s_E$ &-0.034814589   &   0.49412	  (yrs) \\ 
 \label{tab:frefond}
\end{tabular}
\end{center}
\end{table}
}

\def\tableXQP_relat
{
\begin{table*}
{
\begin{tabular}{crrrrrrrr }
$p$   &  Frequency    & $\ab\alpha_p^{(r)}$ & $\beta_p^{(r)}$ & $\ab\alpha_p^{(J)}$ & $\beta_p^{(J)}$& $\ab\alpha_p^{(E)}$ & $\beta_p^{(E)}$& $\varphi_p^{(r)}$ \\
\noalign{\smallskip}
& (rad/day) & (km) & (rad) & (km) & (rad) & (km) & (rad) &  (rad) \\
\hline
\hline
  1 &  -2.14714e-3 &     60.4335555 &      2.5479054 &    13.1270068 &      0.5534404 &     47.3065487 &      1.9944650 &      -1.5835210 \\ 
  3 &  -6.44142e-3 &     20.8015755 &      0.2923293 &      4.5183908 &      0.0634980 &     16.2831847 &      0.2288313 &      -1.6089568 \\ 
  5 &  -1.07357e-2 &     11.0167392 &      0.0929054 &      2.3929886 &      0.0201803 &      8.6237506 &      0.0727251 &      -1.6344092 \\ 
  7 &  -1.50300e-2 &       7.0984885 &      0.0427538 &      1.5418902 &      0.0092867 &      5.5565983 &      0.0334671 &      -1.6609244 \\ 
  9 &  -1.93242e-2 &       4.9877777 &      0.0233653 &      1.0834145 &      0.0050753 &      3.9043632 &      0.0182900 &      -1.6866709 \\ 
 11 &  -2.36185e-2 &      3.6827072 &      0.0141150 &      0.7999351 &      0.0030660 &      2.8827721 &      0.0110490 &      -1.7124149 \\ 
 13 &  -2.79128e-2 &      2.8083507 &      0.0091078 &      0.6100127 &      0.0019783 &      2.1983379 &      0.0071295 &      -1.7381559 \\ 
 15 &  -3.22071e-2 &      2.1908244 &      0.0061577 &      0.4758775 &      0.0013375 &      1.7149470 &      0.0048202 &      -1.7638937 \\ 
 17 &  -3.65014e-2 &      1.7381554 &      0.0043107 &      0.3775515 &      0.0009363 &      1.3606040 &      0.0033743 &      -1.7896277 \\ 
 19 &  -4.07956e-2 &      1.3970712 &      0.0031001 &      0.3034633 &      0.0006734 &      1.0936080 &      0.0024267 &      -1.8153577 \\ 
 21 &  -4.50899e-2 &      1.1345829 &      0.0022778 &      0.2464472 &      0.0004948 &      0.8881358 &      0.0017831 &      -1.8410832 \\ 
 23 &  -4.93842e-2 &      0.9291897 &      0.0017033 &      0.2018329 &      0.0003700 &      0.7273568 &      0.0013333 &      -1.8668040 \\ 
 25 &  -5.36785e-2 &      0.7663012 &      0.0012923 &      0.1664513 &      0.0002807 &      0.5998499 &      0.0010116 &      -1.8925195 \\ 
 27 &  -5.79727e-2 &      0.6356894 &      0.0009926 &      0.1380806 &      0.0002156 &      0.4976089 &      0.0007770 &      -1.9182293 \\ 
 29 &  -6.22670e-2 &      0.5299910 &      0.0007705 &      0.1151214 &      0.0001674 &      0.4148696 &      0.0006031 &      -1.9439331 \\ 
 31 &  -6.65613e-2 &      0.4437859 &      0.0006035 &      0.0963965 &      0.0001311 &      0.3473895 &      0.0004725 &      -1.9696303 \\ 
 33 &  -7.08556e-2 &      0.3730099 &      0.0004765 &      0.0810229 &      0.0001035 &      0.2919870 &      0.0003730 &      -1.9953207 \\ 
 35 &  -7.51499e-2 &      0.3145665 &      0.0003789 &      0.0683282 &      0.0000823 &      0.2462383 &      0.0002966 &      -2.0210036 \\ 
 37 &  -7.94441e-2 &      0.2660642 &      0.0003032 &      0.0577928 &      0.0000659 &      0.2082713 &      0.0002373 &      -2.0466787 \\ 
 39 &  -8.37384e-2 &      0.2256343 &      0.0002439 &      0.0490109 &      0.0000530 &      0.1766234 &      0.0001909 &      -2.0723454 \\ 
 41 &  -8.80327e-2 &      0.1918019 &      0.0001972 &      0.0416620 &      0.0000428 &      0.1501398 &      0.0001544 &      -2.0980031 \\ 
 43 &  -9.23270e-2 &      0.1633917 &      0.0001602 &      0.0354909 &      0.0000348 &      0.1279007 &      0.0001254 &      -2.1236514 \\ 
 45 &  -9.66212e-2 &      0.1394605 &      0.0001307 &      0.0302927 &      0.0000284 &      0.1091677 &      0.0001023 &      -2.1492896 \\ 
 47 &  -1.00916e-1 &      0.1192455 &      0.0001070 &      0.0259018 &      0.0000232 &      0.0933437 &      0.0000837 &      -2.1749173 \\ 
 49 &  -1.05210e-1 &      0.1021264 &      0.0000879 &      0.0221833 &      0.0000191 &      0.0799432 &      0.0000688 &      -2.2005337 \\ 
 51 &  -1.09504e-1 &      0.0875956 &      0.0000724 &      0.0190270 &      0.0000157 &      0.0685686 &      0.0000567 &      -2.2261383 \\ 
 53 &  -1.13805e-1 &      0.0752358 &      0.0000598 &      0.0163423 &      0.0000130 &      0.0588936 &      0.0000468 &      -2.2517303 \\ 
 55 &  -1.18093e-1 &      0.0647024 &      0.0000496 &      0.0140543 &      0.0000108 &      0.0506481 &      0.0000388 &      -2.2773095 \\ 
\end{tabular}
}
\caption{Periodic approximation of the semi-major axes of Janus and Epimetheus. The relative quantities  $a_r$ and $\lam_r$ according to Equations  (\ref{eq:arelat}) and (\ref{eq:larelat}) are represented in the third and fourth columns. The fifth and sixth columns contain the coefficients of $a^{(J)}$ and $\lam^{(J)}$ appearing in Equations (\ref{eq:approx_a}) and  (\ref{eq:approx_la}), while the seventh and eighth columns correspond to the same quantities for Epimetheus.  The frequencies listed in the second column are equal to $p\nu$ where $p$ is in the first column and $\nu$ is given in Table \ref{tab:frefond}. The phases in the ninth column apply directly to Equation (\ref{eq:arelat}) and (\ref{eq:larelat}), and can be adapted to Equation (\ref{eq:approx_a}) and (\ref{eq:approx_la}) by means of Equation (\ref{eq:ratio}).
The origin of time in formulas~(\ref{eq:arelat}) is equal to 1949-Dec-28 00:00:00.0000 (JD 2433278.5) from the ephemeris {\it{Horizons}}.
\label{tab:QP_relat}
}
\end{table*}
}

\def\tableXbulk
{
\begin{table*}[htdp]
\caption{Physical properties of Janus and Epimetheus. (1) Tiscareno \etal (2009). }
\begin{center}
\begin{tabular}{ccc}
 &Janus  & Epimetheus \\
 \hline
 \hline
Mean Radius $^{(1)}$(km)	& 89.5  $\pm$ 1.4 & 58.1 $\pm$ 1.8 \\
Long axis$^{(1)}$	(km) & 101.5 $\pm$ 1.9& 64.9 $\pm$ 2.0\\
Intermediate axis$^{(1)}$	(km) & 92.5 $\pm$ 1.2 & 57.0 $\pm$  3.7\\
Small axis$^{(1)}$	(km) & 76.3 $\pm$  1.2 & 53.1 $\pm$  0.7  \\ 
Triaxiality$^{(1)}$ ${(B-A)}/{C}$&  0.100 $\pm$  0.012 & 0.296$^{+0.019}_{-0.027}$\\
 $\sigb$  (rad/day) &4.95508166  &8.52504483 \\
\end{tabular}
\end{center}
\label{table:bulk}
\end{table*}
}

\def\tableXlibjanus
{
\begin{table}[htdp]
\caption{Frequency analysis in the short-period librational motion of Janus.  This table displays the $7$ first terms of the quasiperiodic decomposition of $\tilde\gamma(t) = e(0)/e(t) \gamma_s(t)$. See the text for more details. }

\begin{center}
{\small
\begin{tabular}{rrrcrrrrrr}
Amp num & Amp ana & Freq     \,\,   \,\,         & $\nc_J$ & $\nu$ &  phase  num \\
(rad)  \,\,        &  (rad)  \,\,        &   (rad/days)&                       &            &   (rad) \\
\hline
\hline
      0.0052021 &       0.0053573 &       9.0109765   &   1 &   0 &        4.705060 \\ 
      0.0014761 &       0.0015412 &       9.0131237   &   1 &  -1 &        3.151328 \\ 
      0.0014756 &       0.0015433 &       9.0088293   &   1 &   1 &        3.117592 \\ 
      0.0002567 &       0.0002159 &       9.0152709   &   1 &  -2 &        1.596407 \\ 
      0.0002521 &       0.0002165 &       9.0066821   &   1 &   2 &        1.532845 \\ 
      0.0001427 &       0.0000201 &       9.0045349   &   1 &   3 &        3.087729 \\ 
      0.0001412 &       0.0000200 &       9.0174181   &   1 &  -3 &        3.183252 \\ 
\end{tabular}
}
\end{center}
\label{tab:libjanus}
\end{table}
}

\def\tableXlibepim
{
\begin{table}[htdp]
\caption{Frequency analysis in the short-period librational motion of Epimetheus.  This table displays the $13$ first terms of the quasiperiodic decomposition of $\tilde\gamma(t) = e(0)/e(t) \gamma_s(t)$. See the text for more details.}

\begin{center}
{\small 
\begin{tabular}{rrrcrrr}
Amp num & Amp ana & Freq     \,\,   \,\,         & $\nc_E$ & $\nu$ &  phase num  \\
(rad)  \,\,        &  (rad)  \,\,        &   (rad/days)&                       &            &   (rad) \\
\hline
\hline
      0.0816497 &       0.0892555 &       9.0088248   &   1 &   1 &        1.356172 \\ 
      0.0813811 &       0.0884486 &       9.0131192   &   1 &  -1 &        1.389845 \\ 
      0.0618972 &       0.0536462 &       9.0152664   &   1 &  -2 &        2.977836 \\ 
      0.0617296 &       0.0546295 &       9.0066776   &   1 &   2 &        2.908991 \\ 
      0.0405482 &       0.0349633 &       9.0109720   &   1 &   0 &        6.085400 \\ 
      0.0172653 &       0.0194522 &       9.0174136   &   1 &  -3 &        4.566668 \\ 
      0.0168758 &       0.0199895 &       9.0045305   &   1 &   3 &        4.462410 \\ 
      0.0073675 &       0.0052782 &       9.0023833   &   1 &   4 &        2.877236 \\ 
      0.0072106 &       0.0050899 &       9.0195608   &   1 &  -4 &        3.006457 \\ 
      0.0052737 &       0.0010462 &       9.0217080   &   1 &  -5 &        4.596183 \\ 
      0.0051221 &       0.0010948 &       9.0002361   &   1 &   5 &        4.429255 \\ 
      0.0024037 &       0.0001873 &       8.9980889   &   1 &   6 &        2.845404 \\ 
      0.0023006 &       0.0001774 &       9.0238552   &   1 &  -6 &        3.038194 \\ 
 \end{tabular}
 }
\end{center}
\label{tab:libepim}
\end{table}
}

\def\tableXJJanus
{
\begin{table}[!ht]
\caption{Numerical values of the coefficients $J_q(\beta_p)$ in the case of Janus. The "$-$" symbol indicates that the corresponding value is lower than $1.\times 10^{-6}$.}
\begin{center}
{\scriptsize
\begin{tabular}{ccccccc}
 $p$&   $J_0(\beta_p)$&   $J_1(\beta_p)$&   $J_2(\beta_p)$&   $J_3(\beta_p)$&   $J_4(\beta_p)$&   $J_5(\beta_p)$  \cr 
\hline
\hline
 1&   0.92488&   0.26626&   0.03732&   0.00346&   0.00024&    0.00001  \cr 
 3&   0.99899&   0.03173&   0.00050&   -&   -&    -  \cr 
 5&   0.99990&   0.01009&   0.00005&   -&   -&    - \cr 
 7&   0.99998&   0.00464&   0.00001&   -&   -&    -  \cr 
 9&   0.99999&   0.00254&   -&   -&   -&    -  \cr 
 11&   1.00000&   0.00153&   -&   -&   -&    -  \cr 
 13&   1.00000&   0.00099&   -&   -&   -&    -  \cr 
 15&   1.00000&   0.00067&   -&   -&   -&    -  \cr 
 17&   1.00000&   0.00047&   -&   -&   -&    -  \cr 
19&   1.00000&   0.00034&   -&   -&   -&    -  \cr 
21&   1.00000&   0.00025&   -&   -&   -&    -  \cr 
23&   1.00000&   0.00018&   -&   -&   -&    -  \cr 
25&   1.00000&   0.00014&   -&   -&   -&    -  \cr 
27&   1.00000&   0.00011&   -&   -&   -&    -  \cr 
29&   1.00000&   0.00008&   -&   -&   -&    -  \cr 
31&   1.00000&   0.00007&   -&   -&   -&    -  \cr 
33&   1.00000&   0.00005&   -&   -&   -&    -  \cr 
35&   1.00000&   0.00004&   -&   -&   -&    -  \cr 
37&   1.00000&   0.00003&   -&   -&   -&    -  \cr 
39&   1.00000&   0.00003&   -&   -&   -&    -  \cr 
\end{tabular}
}
\end{center}
\label{tab:JJanus}
\end{table}
}

\def\tableXJEpim
{
\begin{table}[!ht]
\caption{Numerical values of the coefficients $J_q(\beta_p)$ in the case of Epimetheus}
\begin{center}
{\scriptsize
\begin{tabular}{ccccccc}
 $p$&   $J_0(\beta_p)$&   $J_1(\beta_p)$&   $J_2(\beta_p)$&   $J_3(\beta_p)$&   $J_4(\beta_p)$&   $J_5(\beta_p)$  \cr 
\hline
\hline
 1&   0.22708&   0.57708&   0.35159&   0.12806&   0.03366&    0.00695  \cr 
 3&   0.98695&   0.11367&   0.00652&   0.00025&   -&    -  \cr 
 5&   0.99868&   0.03634&   0.00066&   -&   -&    -  \cr 
 7&   0.99972&   0.01673&   0.00014&   -&   -&    -  \cr 
 9&   0.99992&   0.00914&   0.00004&   -&   -&    -  \cr 
 11&   0.99997&   0.00552&   0.00002&   -&   -&    -  \cr 
 13&   0.99999&   0.00356&   -&   -&   -&    -  \cr 
 15&   0.99999&   0.00241&   -&   -&   -&    -  \cr 
 17&   1.00000&   0.00169&   -&   -&   -&    -  \cr 
19&   1.00000&   0.00121&   -&   -&   -&    -  \cr 
21&   1.00000&   0.00089&   -&   -&   -&    -  \cr 
23&   1.00000&   0.00067&   -&   -&   -&    -  \cr 
25&   1.00000&   0.00051&   -&   -&   -&    -  \cr 
27&   1.00000&   0.00039&   -&   -&   -&    -  \cr 
29&   1.00000&   0.00030&   -&   -&   -&    -  \cr 
31&   1.00000&   0.00024&   -&   -&   -&    -  \cr 
33&   1.00000&   0.00019&   -&   -&   -&    -  \cr 
35&   1.00000&   0.00015&   -&   -&   -&    -  \cr 
37&   1.00000&   0.00012&   -&   -&   -&    -  \cr 
39&   1.00000&   0.00010&   -&   -&   -&    -  \cr 
\end{tabular}
}
\end{center}
\label{tab:JEpim}
\end{table}
}

\def\tableXdefvar
{
\begin{table}[!ht]
\caption{Definition of main notations used in the paper }
\begin{center}
{\scriptsize
\begin{tabular}{rcl}
\hline
\hline
$ \mu$ & = & Constant of the third Kepler law including $J_2$ (see Formula (\ref{eq:kepler}))\\
$ J_2$ & = & Oblateness of Saturn  \\
$ m_i$ & = & Mass of the body where the subscript $i$ characterized the body\\
$ M_S$ & = & Mass of Saturn\\
$ \zeta_i$ & = & reduced mass ($m_i/(m_J + m_E)$) \\
$ R_i$ & = & Radius of the body where the subscript $i$ characterized the body\\
$[I]$ & = & the normalized inertia tensor. \\
$A,B,C$ & = & Normalized ($m_i R_i^2$) moments of inertia of the whole satellites  \\
               &   &  $(A<B<C)$. \\
$\theta$ & = & Rotation angle defined as the angle between the orientation\\   
& &of the principal axes $A$ and the line of node of the orbit. \\
$\gamma$ & =& physical libration, oscillation around a mean uniform motion.\\
$ \ell $ & = & Mean Anomaly \\
$ f $ & = & True Anomaly \\
$\lambda$ & =& mean longitude of the orbit \\
$\lambda_r$ & =& relative mean longitude ($\lambda_J - \lambda_E$)\\
$\beta_i$ & =& Amplitude of the expanding series of $\lambda$ \\
$v$ & =& draconic true longitude of the orbit  $(f+\omega)$ \\
$\omega$ & =& argument of pericenter \\
$\varpi$ & =& longitude of pericenter  \\
$\Omega$ & =& longitude of ascending node \\
$a$ & =& semi-major axis. \\
$\bar{a}$ & =& mean semi-major axis \\
$a_r$ & =& relative semi-major axis ($a_J - a_E$)\\
$\alpha_i$ & =& Amplitude of the expanding series of $a$ \\
$e$ & =& eccentricity. \\
$\bar{e}$ & =& mean eccentricity  \\
$g$ & =& frequency of the longitude of pericenter ($\varpi$). \\
$s$ & =& frequency of the node ($\Omega$). \\
$\bar{n}$ & =& mean mean motion ($\lambda$) \\
$\nc$ & =& $\nb - g$  \\
$\nu$ & =& orbital libration frequency. \\
$\sigma$ & =& frequency of the free libration. \\
$\bar{\sigma}$ & =& mean frequency of the free libration \\
\end{tabular}
}
\end{center}
\label{tab:defvar}
\end{table}
}

\begin{document}                                                                   

\title{Analytical description of physical librations of Saturnian coorbital satellites Janus and Epimetheus}
\author{Philippe Robutel$^{\text{ a}}$, Nicolas Rambaux$^{\text{ b, a}}$, Julie Castillo-Rogez$^{\text{ c}}$\\ \\
\it
$^{\text{ a}}$ ASD, IMCCE-CNRS UMR8028,
Observatoire de Paris, Paris, France\\
$^{\text{ b}}$ Universit\'e Pierre et Marie Curie - Paris 6, France\\
$^{\text{ c}}$  Jet Propulsion Laboratory, Caltech, Pasadena, USA\\
 }

\date{\today}
\maketitle

\begin{abstract}

Janus and Epimetheus are famously known for their distinctive horseshoe-shaped orbits resulting from a 1:1 orbital resonance. Every four years these two satellites swap their orbits by a few tens of kilometers as a result of their close encounter. Recently \citet{TiThBu2009} have proposed a model of rotation based on images from the Cassini orbiter. 
These authors inferred the amplitude of rotational librational motion in longitude at the orbital period by fitting a shape model to the recent Cassini ISS images. By a quasiperiodic approximation of the orbital motion, we describe how the orbital swap impacts the rotation of the satellites. To that purpose, we have developed a formalism based on quasi-periodic series with long and short-period librations. In this framework, the amplitude of the libration at the orbital period is found proportional to a term accounting for the orbital swap. We checked the analytical quasi-periodic development by performing a numerical simulation and find both results in good agreement. To complete this study, the results regarding the short-period librations are studied with the help of an adiabatic-like approach.

\end{abstract}

\section{Introduction}
\label{sec:intro}
The orbital motion of Janus and Epimetheus presents a peculiar horseshoe-shaped orbit resulting from a 1:1 orbital resonance (e.g. \citealt{DeMu1981b}; \citealt{YoCoSyYo1983} ; \citealt{MuDe1999}; \citealt{JaSpPoBeCoEvMu2008} and references therein). Every four years the two satellites swap their orbits by a few tens of kilometers as a result of their close encounter. As the mass of Janus is 3.6 times greater than the mass of Epimetheus, the dynamical motion of the latter is more sensitive to the swap than the dynamical motion of Janus.

The rotational motion of the satellites depends mainly on the gravitational torque of Saturn acting on the dynamical figure of each moon. The expression of the gravitational torque is: 
\begin{equation}
\vec{T} = \frac{3GM_S}{r^3}  \vec{u} \times [I] \vec{u}
\label{torquepointless}
\end{equation}
with $G$ the gravitational constant, $M_S$ the mass of Saturn, $[I]$ the  inertia tensor of the moon, $r$ the distance between Saturn and the moon, and $\vec{u}$ the unit vector toward Saturn in the moon's reference frame. The gravitational torque $\vec T$ depends on the relative Saturn-moon distance, hence the swap also yields his signature on the rotational motion of the satellites. 

First estimates of the rotational motion of the two coorbital satellites Janus and Epimetheus have been obtained by \citet{TiThBu2009}. From images provided by the Cassini orbiter, they fitted a numerical shape model of the moons, which included the amplitude of the libration in longitude. The libration in longitude corresponds to the oscillation of the body along its equatorial plane. \citet{TiThBu2009} obtained an amplitude of $5.9^{\circ} \pm 1.2^{\circ}$ for Epimetheus. For Janus, the uncertainty on the fit of the libration determination is too large to yield an accurate librational amplitude. However, based on their shape model, Tiscareno et al. suggested a value of $0.33^{\circ} \pm 0.06^{\circ}$ for the amplitude of the libration in longitude. In addition, they identified an unexplained constant phase of  $5.3^{\circ} \pm 1^{\circ}$ for Janus, whereas for Epimetheus such offset is in the error bar.
A recent numerical study by \cite{No2010} explored the three-dimensional rotational motion of these satellites based on the numerical shape deduced by \citet{TiThBu2009}. Noyelles' study suggests a strong influence of the swap on the rotational motion of Janus and Epimetheus, which we propose to explore in the present study. 

The orbital motion of Janus and Epimetheus appears to be very regular, at least over a timescale of several thousands of years. Thus, the trajectories of these satellites can be considered as quasi-periodic.   Schematically these trajectories evolve on three different timescales. The shortest corresponds to the mean motion of the satellites with a period of about $0.7$ days. The second component has a period of $8$ years and is associated with the close encounters of the satellites. The long-period component is the secular variations of the satellites eccentricities and inclinations over periods of a few thousands years. Since the rotation of Janus and Epimetheus is synchronous with their orbital motion, it reflects these different timescales.

One of the goals of this paper is to understand the influence of the $8$-year horseshoe motion on the rotational librations of Janus and Epimetheus. To this purpose, we develop an analytical solution of the rotation of these bodies that can simulate the main features of their spin. 
Then, in Section \ref{sec:orbit}, we model the orbits of the two satellites through quasi-periodic expansions. The fourth section is dedicated to the description of the librational motion of each satellite.
We develop three approaches to describe the librational motion in details: (1) a quasi-periodic development that highlight the fundamental frequencies involved in this problem, (2) an adiabatic invariant approach that focuses on the short-period librations, and (3) a numerical approach to reach high accuracy. 
Then we discuss the offset of Janus's orientation and investigate the effect of high spherical harmonics (order 3) and tidal coupling. We also discuss the influence of triaxiality on libration amplitudes.

\section{Physical librations for Keplerian orbit}
\label{sec:KO}
First, let us recall the librational response of a satellite in synchronous spin-orbit resonance with a fixed Keplerian orbit (constant semi-major axis $a$ and eccentricity $e$). 
The position of the moon is determined through its relative distance with Saturn $r$, and its orbital longitude is defined by the draconic true longitude (angle between the body and its line of nodes) denoted by $v$. The mean anomaly will be denoted by $\ell$, while the true anomaly is denoted by $f$. We neglect the effect of the obliquity, which is small \citep{No2010}, so that the  orientation of the body is specified by the angle $\theta$ defined with respect to the line of nodes.  

The dynamical equation governing the rotation of the moon is the angular momentum balance with the gravitational torque exerted by Saturn.  If we introduce the physical libration $\gamma$   by the relation $\theta = \ell + \omega + \gamma$, where the angle $\gamma$  represents the oscillations around the uniform synchronized rotation and $\omega$  the fixed argument of the pericenter, the angular momentum equation reads:  
\be
 \ddot \gamma  + \frac{\sigma^2}{2} \left(\frac{a}{r}\right)^3 \sin2( \gamma + \ell -f) = 0, 
\label{gamma}
\ee

The frequency $\sigma$ is the frequency of the free libration, also called frequency of the proper libation. It is equal to $\sigma = n \sqrt{3 (B-A)/C}$ where $A<B<C$ are the normalized moments of inertia of the satellite and $n$ its mean motion. 

For a small eccentricity $e$, the difference between the true and the mean anomalies is approximated at first-order in $e$ by $f - \ell = 2 e \sin{\ell}$. In addition, we approximate $a = r$, and, for $\gamma$ small, the linearized equation of Eq.~(\ref{gamma}) is 
\be
 \ddot \gamma   + \sigma^2 \gamma = 2 e \sigma^2 \sin{\ell} 
\label{gamma_approx}
\ee

 The librational solution is then simply
\be
\gamma =  A_{\gamma} \sin{(\sigma t + \phi_{\gamma})} + \frac{2e \sigma^2 }{\sigma^2 - n^2}  \sin \ell 
\label{sol:gammaK}
\ee 
where $A_{\gamma},  \phi_{\gamma}$ depend on the initial conditions and the right-hand side term is the forced libration. The forced libration oscillates at the mean motion frequency, and its amplitude is proportional to the ratio of the eccentricity to the difference between the square of the libration frequency $\sigma$ and the square of the forced frequency $n$. Therefore, the amplitude of the forced libration depends on both the magnitude of the torque and the proximity of the free libration frequency to the orbital frequency. 

In the following sections, we describe the orbits and investigate the impact of the horseshoe-shaped orbit on the physical librations.

\section{Orbital description of Janus and Epimetheus}
\label{sec:orbit}

\subsection{Osculating elliptical elements and fundamental frequencies}
\label{sec:osc}

The co-orbital satellites Janus and Epimetheus are famously known to exchange their orbits every four years. This swap takes a short time-span, which does not exceed six months. In order to model the peculiar orbital motion of these satellites, we numerically integrate the three-body problem composed of Saturn, Janus, and Epimetheus, including the oblateness $J_2$ of Saturn. By using the frequency analysis developed by \citet{Laskar1988,Laskar2005} for the purpose of Celestial Mechanics studies, we express the numerical solution as a quasi-periodic function expanded in Fourier series, where each frequency is a linear combination (with integer coefficients) of six fundamental frequencies (proper frequencies) denoted by $(\nb,\nu,g_J,g_E,s_J,s_E)$ (in the following, we use the subscript  J for Janus and E for Epimetheus). The first of these frequencies, $\nb$, called proper mean motion, is associated to the mean orbital motion common to the two co-orbital satellites, and it is constant along satellite orbits.  In the same way, the five other proper frequencies are also constant and might be considered as integrals of movement \citep{Laskar2005}. The second fundamental frequency $\nu$ corresponds to the  libration frequency along the horseshoe orbits, while the four last ones are associated to the motion of the pericenters ($g_E,g_J$) and of the ascending nodes ($s_E,s_J$).  

If the gravitational interactions between Janus and Epimetheus were negligible, then the precessions of the pericenters and of the nodes, which are only due to $J_2$, would satisfy the relations $g_J = g_E$ and $s_ J = s_E$, and the eccentricities and inclinations of the two satellites would be constant (excepted for small short-period variations). However, due to the satellites mutual interactions, the proper frequencies of precessions of the orbits are slightly different, and both eccentricities and inclinations undergo large long-term variations whose periods are about $2760$ and $7690$ years, respectively. The long-term variations of the eccentricities will be neglected in the present work. Their impact on the rotation of the co-orbital satellites is mentioned in section \ref{sec:num_QP}. 

The initial conditions of the numerical integrations come from the ephemeris {\it{Horizons}} \citep{Gioetal1996}. The model used in the ephemeris {\it{Horizons}} takes into account the gravitational interactions with the Sun and the other main satellites of the Saturnian system \citep{JaSpPoBeCoEvMu2008}. Thus, in order to be consistent with that model, we fit the initial conditions of our integration in such a way that the fundamental frequencies $\nb$ and $\nu$ are the same in both cases. The entire set of fundamental frequencies, as well as their associated periods, are displayed in Table \ref{tab:frefond}. The frequencies $(g_J, g_E)$ and $(s_J, s_E)$ are very closed indicating that the mutual interactions are small.
 
 \prep{
\tableXfrefond
}

 Due to the high orbital precession rate generated by Saturn's oblateness, we define the elliptic elements by introducing Saturn's $J_2$ in the third Kepler's law, which yields:
\icar{
\be
n^2a^3 = \mu, \quad \text{with } \quad \mu = G(M_S + m_J + m_E)\left(1 + \frac{3}{2}J_2\left(\frac {R_S} \ab\right)^2\right)
\label{eq:kepler}
\ee
}
\prep{
\be
\begin{split}
& n^2a^3  = \mu, \quad \text{with } \cr
&\mu        = G(M_S + m_J + m_E)\left(1 + \frac{3}{2}J_2\left(\frac {R_S} \ab\right)^2\right)
\end{split}
\label{eq:kepler}
\ee
}

where $M_S$, $R_S$ are the mass and equatorial radius of Saturn, $m_J, m_E$ the masses of Janus and Epimetheus, and $\ab$ the    barycenter of semi-major axis Janus and Epimetheus (see Formula (\ref{eq:approx-a})). 
The elliptic elements $(a,e,I,\lam,\varpi,\Omega)$\footnote{The index "J" or "E" is added to specify that the elements are related to Janus or Epimetheus if necessary.} are the elements of the unique ellipse tangent to the planetocentric velocity  at the planetocentric location of the satellite, assuming that the third Kepler law is given by equation (\ref{eq:kepler}).
 As shown by Figure~\ref{fig:a}, this definition of elliptic elements removes the main orbital oscillations from the elliptical elements \citep{Gr1981}. In addition, it is well-known that changing the value of $\mu$ shifts the mean value of the semi-major axis by a quantity of the order of $J_2 (R_S/\ab)^2$, which leads to the translation of about $600$ km clearly visible in Fig.
\ref{fig:a}.
\prep{
\figureXa
}

\subsection{Analytical expression of the elliptic elements' variations}
\label{sec:orb_ana}

In this section, we detail the quasi-periodic expansion of the elliptic elements for both satellites that will be useful for the rotation study.

According to classical theories (i.e \citealt{DeMu1981b,YoCoSyYo1983} or \citealt{Namouni1999}, for more recent developments),  the variations of the mean longitudes and semi-major axes of the co-orbital satellites are accurately approximated by the expressions:  
\icar{
\begin{align}
&\lambda_J \approx \lam_0 + \nb t  + \zeta_E \lam_r \,\,  {\rm mod}(2\pi),  & \lambda_E \approx \lam_0+ \nb t  - \zeta_J \lam_r \,\,  {\rm mod}(2\pi)  \label{eq:approx-la} \\ 
&a_J \approx {\bar a }   + \zeta_E a_r ,  &  a_E \approx {\bar a }   - \zeta_J a_r\phantom{\nb t  - \zeta_J \lam_r \,\,  {\rm mod}(2   }   \label{eq:approx-a}
\end{align}
}
\prep{
\be
\begin{split}
&\lambda_J \approx \lam_0 + \nb t  + \zeta_E \lam_r \,\,  {\rm mod}(2\pi)  \cr
& \lambda_E \approx \lam_0 + \nb t  - \zeta_J \lam_r \,\,  {\rm mod}(2\pi) 
\end{split}
 \label{eq:approx-la} 
\ee
\be
\begin{split}
&a_J \approx {\bar a }   + \zeta_E a_r   \\
&  a_E \approx {\bar a }   - \zeta_J a_r\phantom{\nb t  - \zeta_J \lam_r \,\,  {\rm mod}(2   }   
\end{split}
\label{eq:approx-a}
\ee
}

where $ \zeta_J = m_J/(m_J+m_E)$ and  $\zeta_E = 1-\zeta_J$.  The variables $a_r$ and $\lam_r$ represent the relative semi-major axis and mean longitudes of the satellites, that is: $a_r = a_J-a_E$ and $\lam_r = \lam_J - \lam_E$.   These relations clearly reflect the symmetries between the orbits of Janus and Epimetheus. 
Formula~(\ref{eq:approx-a})  implies that the barycenter of the semi-major axes, namely $\zeta_J a_J + \zeta_E a_E$,  is almost constant\footnote{Once averaged the Hamiltonian of the three-body problem on the mean longitude of the satellites, the quantity $\zeta_J\sqrt{a_J} + \zeta_E \sqrt{a_E}$  becomes an  integral of the motion. The relative semi-major axis $a_r$ remaining always very small with respect to $a_J$ and $a_E$, the stated property holds.}. Numerical simulation shows that the relative variation of this quantity is smaller than $2\times 10^{-7}$ if Saturn's oblateness is included in the definition of the elliptic elements, while the variation is about $200$ times more without $J_2$.
Similarly, formula~(\ref{eq:approx-la})  implies that the barycenter of the mean longitudes $\zeta_J \lam_J + \zeta_E \lam_E$ increases almost constantly with time. In other words, $\frac{d}{dt} \left( \zeta_J \lam_J + \zeta_E \lam_E \right) \approx \nb $.
 It turns out that the main variations of $a$ and $\lam$ are  given by the relative motion in coordinates $(a_r, \lam_r)$.

Neglecting the terms of powers greater than two in eccentricities (which are very small for  these satellites), the relative motion of the satellites satisfies the differential system\footnote{Equivalent formulations can be found in \cite{SaYo1988} or  \cite{ReSi2004}}: 
\be
\left\{
\begin{array}{ll}
\dot a_r &=  2\eps \nb\ab \ \left( 1 - (2 - 2\cos\lam_r)^{-3/2} \right)\sin\lam_r \\
  \dot\lam_r   &=  {\displaystyle -\frac{3\nb}{2}\frac{a_r}{\ab}}, \quad {\rm with } \quad 
  {\displaystyle  \eps = \frac{m_J+m_E}{M_S+m_J+m_E} }
\end{array}
\right.
\label{eq:relatif}
\ee

The solutions of these equations can be expanded in a Fourier series as:
\icar{
\begin{align}
a_r(t) & = \ab \left[ \sum_{p \geq 1}  \alpha_p^{(r)}\cos\left( p\nu t + \varphi_p^{(r)}\right)  \right],   \, \text{ with} \quad   \alpha_{2p}^{(r)} = 0 \label{eq:arelat} \\ 
\lam_r(t) & = \pi  +  \sum_{p \geq 1}   \beta_p^{(r)}\sin\left( p\nu t + \varphi_p^{(r)}\right) ,   \, \text{ with} \quad   \beta_{2p}^{(r)} = 0 \label{eq:larelat} 
\end{align}
}
\prep{
\be
\begin{split}
& a_r(t)  = \ab \left[  \sum_{p \geq 1}  \alpha_p^{(r)}\cos\left( p\nu t + \varphi_p^{(r)}\right)  \right] \\
& \text{ with} \quad   \alpha_{2p}^{(r)} = 0 
\label{eq:arelat} 
\end{split}
\ee
\be
\begin{split}
&\lam_r(t)  = \pi  +  \sum_{p \geq 1}   \beta_p^{(r)}\sin\left( p\nu t + \varphi_p^{(r)}\right) \\
 & \text{ with} \quad   \beta_{2p}^{(r)} = 0 
 \label{eq:larelat} 
\end{split}
\ee
}
 where $\nu$ is the frequency of the relative motion, that is the second fundamental frequency defined in Section \ref{sec:osc}. It corresponds to a period of eight years for Janus and Epimetheus. The vanishing of the even coefficients in the series (\ref{eq:arelat}) and (\ref{eq:larelat}) is due to the symmetries of the system (\ref{eq:relatif}). 
 Indeed, the invariance of the differential system by the transformation  $z  \longmapsto -z$ where $z = (a_r,\lam_r)$, leads to the relations $z(t +T/2) = -z(t)$, where $T = 2\pi/\nu$ is the period of the solution. Applying this relation to the Fourier expansion of $z$ we obtain  $ \alpha_{2p}^{(r)} =  \beta_{2p}^{(r)} = 0$.

While the differential system (\ref{eq:relatif}) is integrable, it is hard to get an analytical expansion of its solutions in the form of a Fourier series. Consequently, the coefficients  $ \alpha_p^{(r)}$, $\beta_p^{(r)} $ and the phases $\varphi_p^{(r)}$ have been inferred from the solutions of the numerical integration. A truncated expression of this expansion is given in Table \ref{tab:QP_relat}.  The comparison between the third and fourth columns, which display  the coefficients $\alpha_p^{(r)}$ and $\beta_p^{(r)}$, respectively, emphasizes the different decreasing speed of theses sequences. While the coefficients of the relative semi-majors axis seem to decrease slowly, the sequence  $\beta_p^{(r)} $ converges more rapidly.  Indeed, Equation (\ref{eq:relatif}) imposes  the coefficients $\beta_p^{(r)} $ to be proportional to $ p^{-1}\alpha_p^{(r)}$.

In order to illustrate the convergence of the series (\ref{eq:arelat}) and (\ref{eq:larelat}) towards the solution of the equation of relative motion (\ref{eq:relatif}), we consider different approximations of these series for which the $N$ first terms are summed. To the relative orbit of the two co-orbital satellites in the plan $(a_r,\lam_r)$ visible in Fig. \ref{fig:QP_relat} (bold curve), is superimposed the approximated orbit (dashed curves) obtained by varying the integer $N$. The ellipse obtained for $N=1$ provides a very crude approximation of the relative orbit, while the approximation generated at $N=30$ starts matching accurately the numerical solution. 
In addition to the central symmetry $z \longmapsto -z$ mentioned above,  Fig. \ref{fig:QP_relat} emphasizes a second symmetry with respect to the axis of coordinates. These symmetries impose relationships between the phases $\varphi_p$, whose description is beyond the scope of this paper. 

\prep{
\figureXrelat
}

\prep{
\tableXQP_relat
}

Finally, from the periodic representations (\ref{eq:arelat}) and (\ref{eq:larelat}) of the quantity $z(t)= (a_r(t), \lam_r(t))$ and according to formulas~(\ref{eq:approx-la}) and~(\ref{eq:approx-a}),  the trigonometric approximations of  the semi-major axis and mean longitude of the satellites read: 
\icar{
\begin{align}
a^{(x)}(t) & = \ab \left[ 1 + \sum_{1\leq  p \leq N}  \alpha_p^{(x)}\cos\left( p\nu t + \varphi_p^{(x)}\right)  \right]  = \ab\left[ 1 + \cA_N^{(x)}\right]   \label{eq:approx_a}  \\ 
\lam^{(x)}(t) & = \lam_0 + \bar n t  +  \sum_{1\leq  p \leq N}  \beta_p^{(x)}\sin\left( p\nu t + \varphi_p^{(x)}\right)  =  \lam_0 + \bar n t +\cB_N^{(x)} \label{eq:approx_la} \\
&\alpha_p^{(x)} = (1-\zeta_x )\alpha^{(r)}_p \ab^{-1}, \quad  \beta_p^{(x)} = (1-\zeta_x )\beta^{(r)}_p, \quad 
 \varphi_p^{(J)} = \varphi_p^{(r)}, \quad \varphi_p^{(E)} = \varphi_p^{(r)} +\pi
 \label{eq:ratio}
\end{align}
}
\prep{
\be
\begin{split}
a^{(x)}(t) & = \ab \left[ 1 + \sum_{1\leq  p \leq N}  \alpha_p^{(x)}\cos\left( p\nu t + \varphi_p^{(x)}\right)  \right]  \\
                 &= \ab\left[ 1 + \cA_N^{(x)}\right]  
\end{split}
 \label{eq:approx_a}  
\ee
\be
\begin{split}
\lam^{(x)}(t) & = \lam_0 + \bar n t  +  \sum_{1\leq  p \leq N}  \beta_p^{(x)}\sin\left( p\nu t + \varphi_p^{(x)}\right)  \\
                     &=  \lam_0 + \bar n t +\cB_N^{(x)}
\end{split}
 \label{eq:approx_la} 
\ee
\be
\begin{split}
&\alpha_p^{(x)} = (1-\zeta_x )\alpha^{(r)}_p , \quad  \beta_p^{(x)} = (1-\zeta_x )\beta^{(r)}_p, \\
 &\varphi_p^{(J)} = \varphi_p^{(r)}, \quad \varphi_p^{(E)} = \varphi_p^{(r)} +\pi
 \label{eq:ratio}
\end{split}
\ee
}
where the index $x$ replaces $J$ and $E$, whether we consider Janus or Epimetheus. The numerical values of the coefficients $\alpha_p^{(x)}$ and $\beta_p^{(x)}$ are reported in the fifth and sixth columns of Table~\ref{tab:QP_relat} for Janus and in the seventh and eighth columns for Epimetheus. 
In addition, the ratios of each $\alpha^{(E)}_p/\alpha^{(J)}_p$ and  $\beta^{(E)}_p/\beta^{(J)}_p$ listed in Table~\ref{tab:QP_relat} are close to $3.6$, in agreement with the formulae~(\ref{eq:ratio}). The short-period oscillations do not appear in that table because they are negligible in comparison to the other parameters.

In the following section, we use this representation of the elliptical elements of the satellites, especially for the mean longitudes, to develop an elementary perturbation theory describing the rotation of Janus and Epimetheus. We consider in the next section that the eccentricities and inclinations are constant, as underlined in section \ref{sec:osc} and discussed in Section \ref{sec:num_QP}. In the same way, the precession of the pericenters and nodes will be approximated by assuming uniform motion defined as: 

\begin{align}
e(t) & = \bar e,   \quad \varpi(t) = gt + \varpi_0 \label{eq:approx_e}\\
I(t) & = \bar I,  \quad \Omega(t) = st + \Omega_0 \label{eq:approx_I}
\end{align}

\section{Physical librations for Janus and Epimetheus}
\label{sec:rotation}

\subsection{Perturbative analysis}
\label{sec:pert}
\subsubsection{Dynamical equations}
\label{sec:dyn-eq}
The equation governing the physical libration is inferred from the angular momentum balance equation projected onto the equatorial plane of the body 
\be 
 \ddot{\theta} + \frac32\frac{B-A}{C}\frac{GM_S}{r^3}  \sin{2(\theta - v)} =0 
\label{eq:theta}
\ee
where all variables have been defined in Section 2. Since the angle $\theta - v$ remains always small, the linearization of the equation (\ref{eq:theta}) is a valid approximation to the rotation. Therefore,  in this section, we will consider the linear time-dependent equation:
\be
\ddot{\theta} +\sigma^2\left(\frac{a}{r}\right)^3  (\theta -v) =0, \, \text{with } \,\sigma^2 = 3 \frac{GM_S}{a^3} \frac{(B-A)}{C} 
\label{eq:thetalin} 
\ee 
This equation is not integrable because the quantities $\sigma$, $a/r$ and $v$ are implicit functions of time.  However, by using the expansions in quasi-periodic functions of time presented in section \ref{sec:orbit} for the orbital elements,  and an elementary perturbation theory, we can obtain the approximated solution of this equation.
Let us first introduce the physical libration $\gamma$. This angle is defined as the oscillation of $\theta$ around the uniform motion $(\bar{n} - s)t$, where $\bar{n} - s$ is the main frequency of  the draconic mean longitude $\lambda - \Omega = \ell + \omega$. Therefore the physical libration reads: $ \gamma = \theta - (\bar{n} - s) t - \ell_0 -\omega_0 $, the angles $\ell_0$  and $\omega_0$ being the initial values of the mean anomaly and of the argument of the pericenter of the considered satellite\footnote{The definition of $\gamma$ is different from the one used in the Keplerian case for in the present situation the angle $\ell + \omega$ is not proportional to the time.}.
Consequently, using the relations (\ref{eq:approx_la}), (\ref{eq:approx_e}), (\ref{eq:approx_I}),  the definition of $\gamma$ and $v = f + \omega$ for the true anomaly $f$, the angle $\theta -v$ also reads:
\icar{
\be
\begin{split}
\theta -v &= \gamma - v +(\nb-s)t +\ell_0 +\omega_0 = \gamma - [\ell +\omega - (\bar{n} -s)t - \ell_0 - \omega_0 ] - [f - \ell] \\
              & =   \gamma -\cB_N  - 2e\sin\ell
 \end{split}
 \label{eq:centre}
\ee
}
\prep{
\be
\begin{split}
\theta -v &= \gamma - v +(\nb-s)t +\ell_0 +\omega_0  \\
               & = \gamma - [\ell +\omega - (\bar{n} -s)t - \ell_0 -\omega_0 ] - [f - \ell] \\
              & =   \gamma -\cB_N  - 2e\sin\ell
 \end{split}
 \label{eq:centre}
\ee
}
where terms of order 2 and greater in eccentricity have been neglected.
At this point, it is convenient to use the function $y =  \gamma -\cB_N$.
 Indeed, although the amplitudes of the terms contained in $\cB_N$ are very large, their frequencies are small and therefore the acceleration generated by  $\cB_N$, which is of order $\nu^2$, is negligible with respect to $\sigma^2$.  Under these approximations, Equation (\ref{eq:thetalin}) becomes:
\be
\ddot y + \sigma^2\left(\frac{a}{r}\right)^3 y = 2e\sigma^2\left(\frac{a}{r}\right)^3\sin\ell
\label{eq:gammalin}
\ee
This equation has the same form as in the Keplerian case (see Equation (\ref{gamma_approx})), but it is not periodic anymore but quasiperiodic on the time because, according to formulas~(\ref{eq:approx_la}) and (\ref{eq:approx_e}), $\ell$ reads: 
\be
\ell =  \lambda -\varpi =   \ell_0 + \nc t  +  \cB_N, \quad \text{with }\, \Check{n} = \bar n -g 
\label{eq:ell}
\ee 
so
\icar{
\be
e^{i\ell} = e^{i\ell_0}e^{i\nc t}       \prod_{1\leq  q \leq N} e^{i \beta_q\sin\left( q\nu t + \varphi_q\right)} =
                  e^{i\ell_0}e^{i\nc t}       \prod_{1\leq  q \leq N} \sum_{k\in \ZZ}  J_k( \beta_q)e^{ik\left( q\nu t + \varphi_q\right)}
\label{eq:devel_l} 
\ee
}
\prep{
\be
\begin{split}
e^{i\ell}  &= e^{i\ell_0}e^{i\nc t}       \prod_{1\leq  q \leq N} e^{i \beta_q\sin\left( q\nu t + \varphi_q\right)} \\
               & =
                  e^{i\ell_0}e^{i\nc t}       \prod_{1\leq  q \leq N} \sum_{k\in \ZZ}  J_k( \beta_q)e^{ik\left( q\nu t + \varphi_q\right)}
\end{split}
\label{eq:devel_l} 
\ee
}
where the $J_k( x)$ are the Bessel functions (see  Appendix~\ref{a:bessel}). Let us recall that in the previous expression, the coefficients $\beta_q$ vanish when $q$ is even.  Applying usual properties of the Bessel functions that are recalled in Appendix~\ref{a:bessel}, we deduce that, for $q$ odd, the ratio between the coefficients $J_k( \beta_q)$ computed for Epimetheus and for Janus is well approximated by:
\be
\left\vert \frac{ J_k( \beta_q^{(E)})}{ J_k( \beta_q^{(J)})} \right\vert \approx \left( \frac{\zeta_J}{\zeta_E}\right)^{\vert k\vert}  \approx 3.6^{\vert k\vert}
\ee
For this reason, the coefficients of the expansion (\ref{eq:devel_l}) decrease with increasing $k$ much more rapidly for Janus than for Epimetheus.   Consequently, the number of terms necessary to approximate $e^{i\ell}$ to a given accuracy using a truncated expression of (\ref{eq:devel_l}) is different for the two moons (see Section \ref{sec:compar}). 
In order to simplify the following development, we present the series containing only the first term $N=1$, which corresponds to approximate the relative orbit of the two bodies by the green ellipse in figure \ref{fig:QP_relat}. This makes a crude simplification regarding the orbital motion of the moons but the generalization to N higher than 1 is then straightforward.
Then we have:
\be
e^{i\ell} = e^{i\ell_0}e^{i\nc t}        \sum_{k\in \ZZ}  J_k( \beta_1)e^{ik\left( \nu t + \varphi_1\right)}
\label{eq:expil}
\ee
It turns out that, under these approximations and assuming as in (\ref{eq:approx_e}) that the eccentricity is constant and denoted $\eb$:
\icar{
\be
\begin{split}
e\sin\ell  =  \eb\cS  \\
      \text{with } \cS =  &\phantom{+}   J_0(\beta_1) \sin{( (\nc t + \ell_0)} +  \\
      &    \sum_{p \geq 1}   J_p(\beta_1) \left[(\sin( (\nc +p\nu )t +  p\varphi_1 + \ell_0) + (-1)^p \sin( (\nc -p\nu)t - p\varphi_1 + \ell_0)\right] 
\end{split}
\label{eq:v}
\ee
}
\prep{
\be
\begin{split}
&e\sin\ell  =  \eb\cS  \quad  \text{with } \\
    & \cS =  \phantom{+}   J_0(\beta_1) \sin{( \nc t + \ell_0)} +  \\
      &    \sum_{p \geq 1}   J_p(\beta_1) \big[(\sin( (\nc +p\nu )t +  p\varphi_1 + \ell_0) +  \\
      &  \phantom{\sum_{p \geq 1}   J_p(\beta_1)}   (-1)^p \sin( (\nc -p\nu)t - p\varphi_1 + \ell_0)\big] 
\end{split}
\label{eq:v}
\ee
}
Let us mention that in the expression (\ref{eq:expil}) the index of the summation $k$ belongs to $\ZZ$, while in (\ref{eq:v}) the summation is restricted to positive integers, p.
Using (\ref{eq:expil}) and expanding $(a/r)^{3}$ at first order in eccentricity, we also get the expression of:
\be
\left( \frac{a}{r}\right)^3 =  1+ 3\eb \cC
\label{eq:aSr3}
\ee
The expression of $\cC$ is the same as $\cS$, where sine functions are replaced by cosine functions.
The last term that we have to expand is $\sigma^2 = \sigb^2(\ab/a)^{3}$.  By (\ref{eq:approx_a}) with $N =  1$, we have $a^3 \approx \ab^3\left(  1+3 \cA_1 \right)$ and consequently: 
 \be
\sigma^2  \approx  \sigb^2\left(  1-3 \cA_1 \right)\, \text{ with} \quad \sigb^2 = 3\frac{Gm}{\ab^3}\frac{B-A}{C}
\label{eq:sig}
\ee

By substitution of the relations  (\ref{eq:v}),  (\ref{eq:aSr3}) and (\ref{eq:sig})  in the  expression (\ref{eq:gammalin}) this equation becomes: 
\be
\ddot y + \sigb^2(1-3\cA_1)(1 +3\eb\cC )y  = 2\eb\sigb^2(1-3\cA_1)(1 +3\eb\cC )\cS
\ee
If we split $y$ into a sum of terms of decreasing magnitude as $y = y_0 + y_1 + \cdots $, we obtain the following system of equations:
\begin{align} 
\label{eq:approx1}
&\ddot y_0 + \sigb^2  y_0 = 2\eb\sigb^2\cS   \\
\label{eq:approx2}
& \ddot y_1 + \sigb^2  y_1 =   3\sigb^2\left(\cA_1 - \eb\cC \right )y_0  +6\eb\sigb^2\left( \eb\cC- \cA_1 \right)\cS  \\
& \phantom{\ddot \gamma_1 + \sigma^2  \gamma_1 = \sigma^2} \vdots \nonumber
\end{align}

The solutions of these equations are trigonometric series, whose frequencies are linear combinations with integer coefficients of fundamental frequencies of the satellites $(\nb,\nu,g,s)$ and of the frequency of the free libration  $\sigb$.  The solution independent of the free frequency  $\sigb$ is usually called "forced solution" and it is a quasi-periodic series of frequencies  $(\nb,\nu,g,s)$.

When weak dissipation is introduced in the system, almost trajectories converge towards a quasiperiodic attractor (see  \cite{CeCh2008}). This quasiperiodic trajectory is very close to the forced solution described in the conservative system. Indeed, the fundamental frequencies of these two solutions are the same, only the amplitudes and phases are slightly different (see section \ref{sec:tidal}). 
Consequently, it is relevant to focus on the forced solution in the conservative problem. 

\subsubsection{Forced librations}
\label{sec:forced}
To begin with, let us associate  to a quasi-periodic function $f$ the function  $\hat f$ such that:
\icar{
\be
  \text{ if } \quad f(t) = \sum_p f_p\sin(v_pt+\phi_p) \, \text{ then }\quad \quad {\hat f}(t) = \sum_p \frac{\sigb^2}{\sigb^2 - v_p^2}f_p \sin(v_pt+\phi_p) 
   \nonumber
 \ee
}
\prep{
\be
\begin{split}
  &\text{ if } \quad f(t) = \sum_p f_p\sin(v_pt+\phi_p) \\
  & \text{ then }\quad \quad {\hat f}(t) = \sum_p \frac{\sigb^2}{\sigb^2 - v_p^2}f_p \sin(v_pt+\phi_p) 
\end{split}
   \nonumber
 \ee
}
With these notations, the general solution of  (\ref{eq:approx1}) reads:
\be
y_0(t) = h\sin(\sigb t +\psi) + 2\eb \hat S
\label{eq:sol_0}
\ee
where $h$ and $\psi$  are arbitrary constants. In this section, we focus on the forced solution, so we put $h=0$.
As a consequence, It is easy to verify in (\ref{eq:approx1}) that the contributions of $\eb y_0\cC$ and $\eb^2\cC\cS$, denoted respectively $\eb \widehat{ y_0\cC}$ and $\eb^2 \widehat{\cC\cS}$, is a second-order in eccentricity, and that $ \widehat{ \cA_1 y_0}$ and $\eb \widehat{\cA_1\cS}$  are of order $\eb\alpha_1$. As, for Janus and Epimetheus, the coefficient $\alpha_1$ is lower than $\eb$, then the term $y_1$ can be neglected. Finally, the forced libration can be approximated by the expression:
\icar{
\be
\begin{split}
\gamma(t) =  &\sum_{1\leq  q \leq N}   \beta_q\sin\left( q\nu t + \varphi_q\right) 
                   +\frac{ 2\eb \sigb^2J_0(\beta_1)}{\sigb^2 - \nc^2} \sin{( \nc t + \ell_0)} +  \\
      &   2\eb \sigb^2\sum_{p \geq 1}   J_p(\beta_1) \left[\frac{\sin( (\nc +p\nu )t +  p\varphi_1 + \ell_0)}{\sigb^2 - (\nc +p\nu )^2} - (-1)^p
      \frac{\sin( (\nc -p\nu)t - p\varphi_1 + \ell_0)}{\sigb^2 - (\nc - p\nu )^2}\right]
\label{eq:sol_forced}
\end{split}
\ee
}
\prep{
\be
\begin{split}
&\gamma(t) =  \sum_{1\leq  q \leq N}   \beta_q\sin\left( q\nu t + \varphi_q\right) \\
                   &+\frac{ 2\eb \sigb^2J_0(\beta_1)}{\sigb^2 - \nc^2} \sin{( \nc t + \ell_0)} +  \\
      &   2\eb \sigb^2\sum_{p \geq 1}   J_p(\beta_1) \left[\frac{\sin( (\nc +p\nu )t +  p\varphi_1 + \ell_0)}{\sigb^2 - (\nc +p\nu )^2} \right. -\\
      & \left.  \phantom{\sum_{1\leq p }   J_p(\beta_1)} (-1)^p \frac{\sin( (\nc -p\nu)t - p\varphi_1 + \ell_0)}{\sigb^2 - (\nc - p\nu )^2}\right]
\label{eq:sol_forced}
\end{split}
\ee
}
Let us recall that, in order to give a simple expression of the short-period libration, we put $N=1$ in the expression (\ref{eq:devel_l}). For this reason, only $\beta_1$ appears in the short-period part of  (\ref{eq:sol_forced}). In contrast, in the long-period component of $\gamma$ (first term of Formula (\ref{eq:sol_forced})), $N$ is arbitrary.  The librational angle $\gamma$ is split in two types of terms exhibiting different behaviors. The first type corresponds to the $2\pi/\nu$-periodic terms that depend only on the coefficients $\beta_p$, i.e., on the mean longitudes of the satellites. For these long-period terms, the dynamical figure has no influence. The second type includes terms that vary rapidly (quasi-periodic with short frequencies $\nc \pm p\nu$) and depend on the triaxiality of the body $(B-A)/C$ through the libration proper  frequency $\sigb$. 

The amplitudes of the rapidly oscillating terms depend on the magnitude of the forcing, $2\eb J_p(\beta_1)$ and on the proximity of the forcing frequency $\nc \pm p\nu$ with the libration proper frequency $\sigb$. In the case of Janus, the proper frequency is 4.96 rad/day, which is far from the resonance, whereas for Epimetheus the proper frequency is equal to 8.52 rad/days, and its influence on the amplitude is substantial (see \cite{TiThBu2009}).

By contrast to the Keplerian case, the swap results in the amplitude of the term associated to the frequency $\nc =\nb - g$ to be proportional to $J_0(\beta_1)$. This term is of the order of 1 for Janus, but it is significant in the case of Epimetheus as close to 0.22. Therefore, for both satellites, the rotation significantly departs from the Keplerian case.

\subsubsection{Proper libration}

\label{sec:free}

In this section we investigate the proper libration of the moons (also called free libration) and we especially focus on the influence of a small divisor on the solution. By dissipative effect the proper libration is expected to be small and the damping time scale is short. However excitation mechanism might exist like recent impact for example. To focus on proper libration, let us remove the external forcing by imposing $\eb =0$ in (\ref{eq:approx1}) and (\ref{eq:approx2}). Then, the solution (\ref{eq:sol_0}) reads $y_0(t) = h\sin(\sigb t +\psi)$, where the amplitude $h$ is small but different from zero. Consequently, by substitution of $y_0$ in the equation (\ref{eq:approx2}), we get:
\icar{
\be
\begin{split}
 y_1(t) & =  \frac{3\sigb^2\alp_1h}{2}
  \left(    \frac{\sin((\nu+\sigb)t +\varphi_1+\psi)}{\sigb^2 - (\nu+\sigb)^2}   - \frac{\sin((\nu-\sigb)t + \varphi_1-\psi)}{\sigb^2 - (\nu-\sigb)^2}   \right) \cr
     &  = -\frac32\frac{\sigb\alp_1}{\nu}h \left(      \sin(\nu t+\varphi_1)\cos(\sigb t +\psi)  + O(\frac{\nu}{\sigb})      \right)
\end{split}
\ee
}
\prep{
\be
\begin{split}
 y_1(t) & =  \frac{3\sigb^2\alp_1h}{2}
  \left(    \frac{\sin((\nu+\sigb)t +\varphi_1+\psi)}{\sigb^2 - (\nu+\sigb)^2}   -    \right. \\
  &  \phantom{ y_1  =  \frac{3\sigb^2\alp_1h}{2}} \left. \frac{\sin((\nu-\sigb)t + \varphi_1-\psi)}{\sigb^2 - (\nu-\sigb)^2}   \right) \cr
     &  = -\frac32\frac{\sigb\alp_1}{\nu}h \left(      \sin(\nu t+\varphi_1)\cos(\sigb t +\psi)  + O(\frac{\nu}{\sigb})      \right)
\end{split}
\ee
}
Contrarily to the case  of the forced libration,  the amplitude of the term $\cA_1$, which is equal to  $\alp_1$ (see Table \ref{tab:QP_relat}), plays a major role here. Indeed,  in the present situation,  that term is multiplied by the factor $\sigb/\nu$, which is about $2200$ for Janus and $3700$ for Epimetheus. Thus it generates a second-order solution $y_1$, whose size is comparable to the solution of order one $y_0$.  Then, using the values of $\sigb$ given in Table \ref{table:bulk} for Janus and Epimetheus, the librational responses for the two satellites  are:
\begin{align}
y_J &=  h \left( \sin(\sigb t+\psi) + 0.3 \sin(\nu t+\varphi_1)\cos(\sigb t +\psi) \right) \\
y_E &= h \left( \sin(\sigb t+\psi) + 1.87 \sin(\nu t+\varphi_1)\cos(\sigb t +\psi) \right)
\label{eq:solution}
\end{align}
The proper librations are combinations of a sine term with a constant amplitude and a cosine term with an amplitude varying at the swap frequency. For Janus, the main term is the sine component, whereas for Epimetheus, it is the cosine component.

\subsection{Numerical study and quasiperiodic representation of the libration}
\label{sec:num_QP}

\subsubsection{Quasiperiodic decomposition }
\prep{
\figureXecc
}

\prep{
\figureXerr
}

\prep{
\figureXerrE
}

\prep{
\tableXbulk}

We numerically integrate the orbital and rotational dynamical equations (\ref{eq:theta}) with the triaxialities listed in Table~\ref{table:bulk} in order to determine the accuracy of analytical solution. 
To clearly separate in the frequency analysis the frequencies $(\nc \pm p\nu)$, which are quite close, we have to integrate the trajectories over a long time-span of about $400$ years in the future and $400$ years in the past. We also integrate during $8000$ years to study the very slow variations of the eccentricities of the two satellites. This point will be discussed below.

We focus on the forced libration but the initial conditions of such trajectory are not known. Hence, we use an iterative method based on the frequency analysis to converge towards this trajectory by removing the free libration amplitude (see  \cite{CouLaCoMaUd2010} Section 4.5).

As predicted by the theoretical approach stated in section \ref{sec:forced}, the  libration angle $\gamma=\theta -\nb t -\ell_0 -\omega_0$ can be naturally written as the sum of two components:  $\gamma = \gamma_l + \gamma_s$, where  $\gamma_l$ contains the long-period harmonics and $\gamma_s$  the short-period ones. The frequencies of these harmonics take the form: $\nc \pm p\nu$.
 The long-period oscillations whose amplitude is about $36^\circ$ for Janus and $130^\circ$ for Epimetheus, are very well described by the first summation in the expression (\ref{eq:sol_forced}), that is:
\begin{equation*}
\gamma_l (t)=  \sum_{ q \geq 1, \, \text{$q$ odd}}   \beta_q\sin\left( q\nu t + \varphi_q\right) 
\end{equation*}
  where the numerical values of the coefficients $\beta_q$ appear in the sixth column (for Janus)  and in the eighth one (for Epimetheus) of  Table \ref{tab:QP_relat}. In other words, the Fourier expansion of $\gamma_l$ is similar to the one governing the long-time oscillations of the mean longitude of the corresponding satellite.    
 
 In order to estimate the accuracy of   the coefficients of the short-period terms obtained analytically in Section \ref{sec:forced}, we will focus on  the Fourier decomposition of function $\gamma_s$ deduced from the numerical simulation. 

According to the equation (\ref{eq:gammalin}), the amplitude of the short-period component of the rotational libration $\gamma_s$ is proportional to the eccentricity of the satellite. As this quantity suffers from large very long-period variation (period of about $2760$ years), the amplitude of $\gamma_s$ is slowly time dependent. 
The evolution of the eccentricities are represented in Fig. \ref{fig:ecc}. During the $400$-year numerical integration, the variations of the eccentricities are significant (according to Fig. \ref{fig:ecc}: $e_J$ increases by $\sim20\%$, while $e_E$ decreases by more than $40\%$). Consequently, the amplitudes of the short-period librations, which are proportional to the eccentricities, are modulated with the same ratio. As a consequence, we study the quantity $\tilde\gamma_s(t) = \gamma_s(t) e(0)/e(t)$.    The multiplicative factor $ e(0)/e(t)$, ratio of the initial eccentricity by the eccentricity at the time $t$, imposes to the amplitude of  $\tilde\gamma_s(t)$ to be almost constant during the integration time. 
During the $200$ first years of the simulation, the ratio $e(t)/e(0)$ can be accurately fitted  by the quadratic polynomial  $P$,  equal to $P_J(t) = 1 +1.9839\times10^{ -6}t-4.1467\times10^{-12}t^2$  for Janus, and $P_E(t) = 1 -2.3486\times 10^{-6}t-1.7633\times10^{-12}t^2$ for Epimetheus. The time is counted in Julian days from 1949-Dec-28 00:00:00.0000 (JD 2433278.5). These polynomials approximate the considered ratio with a relative accuracy of $0.32\%$ for Janus and $0.45\%$ for Epimetheus.  Consequently, the value  $\gamma_s(t)$ of the short-period component  libration at an arbitrary time $t$  inside the $200$-years considered interval can easily be deduced from $\tilde\gamma_s(t)$ using the relation $\gamma_s(t) =  P(t) \tilde\gamma_s(t)$.

The main terms of the Fourier decompositions of $\tilde\gamma_s$ are displayed in Table~\ref{tab:libjanus} for Janus and Table~\ref{tab:libepim} for Epimetheus.  The solutions are given in the form:

\icar{
\be
\tilde\gamma(t) = \sum_p \tilde\gamma_p \sin(f_pt +\psi_p) \quad \text{with} \quad f_p =  j_p\nc +k_p\nu
\ee
}
\prep{
\be
\begin{split}
&\tilde\gamma(t) = \sum_p \tilde\gamma_p \sin(f_pt +\psi_p) \\
&  \text{with} \quad f_p =  j_p\nc +k_p\nu
\end{split}
\ee
}
The first and second columns contain the amplitudes of $\tilde \gamma_p$ from the numerical simulation and the analytical expression (\ref{eq:sol_forced}). The comparison between these two columns will be discussed in the next section. The third column contains the frequencies $f_p$, while the integers $j_p$ and $k_p$ are displayed in the fourth and fifth columns. The last column presents the phases $\psi_p$ deduced from the numerical simulation. 

In the case of Janus (Table \ref{tab:libjanus}), the term at the orbital period is dominant, with an amplitude of $0.0052$ radians, that is $0.3^\circ$. By adding the following terms the amplitude is slightly modified. We deduce that, from the polynomial interpolation of $e_J$ given above, the amplitude of $\gamma_s$ increases for $0.34^{\circ}$ in $1950$ to $0.36^{\circ}$ in $2010$, which is very close to the value of $0.33^{\circ}$ given in \cite{TiThBu2009}.

For Epimetheus (Table \ref{tab:libepim}), no dominant term appears clearly and the decrease of the coefficients in slower than for Janus. The amplitude of the short-period oscillations given by the sum of this $13$ terms deceases from $8.77^{\circ}$ in  $1950$ down to $8.35^{\circ}$ in $2010$, which is here again, comparable the $8.9^\circ$ given in \cite{TiThBu2009} (Table~7).

Let us now briefly discuss the accuracy of the quasiperiodic approximations of the numerical solutions.  Due to windowing used in the frequency analysis (see \cite{Laskar2005}), the method does not provide a uniform approximation of studied signal. The accuracy is generally lower on the margin of the considered interval of time. Consequently, the time span of the analysis has been chosen such that the best accuracy is obtained in the interval $I = [0:200]$ years. On the interval $I$, the accuracy is measured with the help of the relative error $\epsilon_p$ defined as:
\be
\epsilon_p = \frac{\underset{t\in I}{\rm Max}\vert \tilde\gamma_s(t) - \tilde\gamma_s^p(t)\vert}{\underset{t\in I}{\rm Max}\vert  \tilde\gamma_s(t)\vert}
\label{eq:error}
\ee
In his expression, $\tilde\gamma_s^p$ is the quasiperiodic approximation of $\tilde\gamma_s$ containing the $p$ dominant terms of the decomposition. As mentioned above, the value of ${\rm Max}\vert  \tilde\gamma_s(t)\vert$ is about $0.34^{\circ}$ for Janus and $8.77^{\circ}$ for Epimetheus.

The number of terms given in Tables \ref{tab:libjanus} and \ref{tab:libepim} is such that the relative error $\epsilon_p$ is better than $10\%$ for both satellites (Janus $8.7\%$, Epimetheus $9.3\%$), which gives an absolute error of about $0.8^\circ$ for Epimetheus and less that $0.03^\circ$ for Janus (The number of selected terms is 13 for Epimetheus and 7 for Janus). The increase in the accuracy with respect to the number of terms included in the quasiperiodic approximation is presented in Fig. \ref{fig:err_coef} where $\log_{10}(\epsilon_p)$ is plotted against $p$. As it was already mentioned, the convergence of the Fourier approximation is much rapid for Janus than for Epimetheus. This figure also shows that  $45$ terms for Janus and $57$ for Epimetheus are necessary to reduce the relative error down to $1.2\%$.

Finally, Fig. \ref{fig:err_Epim} shows the instantaneous relative error $\left( \tilde\gamma_s(t) - \tilde\gamma_s^p(t)\right) / {\rm Max}\vert  \tilde\gamma_s(t)\vert$ for $p=13$ (red dots) and $p=57$ (green dots) in the case of Epimetheus. The results are similar for Janus.
For $p=13$, the discrepancy is mainly due to an over estimate of the influence of the orbital swap on the rotation (one peak every four years). The addition of the harmonics of frequency $\nu \pm p\nc$ for high values of $p$ tends to erase this effect.

\subsubsection{Comparison of the analytical and numerical QP representations }
\label{sec:compar}

The analytical and numerical short-period component of the forced solution are listed in Table  \ref{tab:libjanus} for Janus. 
The accuracy of the short-period terms is a function of the considered harmonic, whose frequency is $\nc \pm p\nu$, where $p$ is an arbitrary positive integer.  
For $p=0$, the discrepancy is about $3\%$ and increases for increasing $p$, e.g. $14\%$ for $p=2$ and to $85\%$ for $p=3$.
This lack of accuracy can be ascribed to neglecting the terms $\beta_q$ for $q$ strictly greater than one in the expansion (\ref{eq:devel_l}), and consequently in the analytical solutions of the rotation (\ref{eq:sol_forced}).  Indeed, as it can be shown by a straightforward calculation, the amplitudes of the terms that have been neglected are given by  $2J_3(\beta_1)J_1(\beta_3)\approx 2\times10^{-4}$ for $p=0$, $J_2(\beta_1)J_1(\beta_3)\approx 4\times10^{-3}$ for $p=1$ and $J_1(\beta_1)J_1(\beta_3)\approx 2\times 10^{-1}$ for $p=2$. These numerical values, which are deduced from Table \ref{tab:JJanus}, are in good agreement with the level of accuracy mentioned above. 

\prep{
\tableXlibjanus
}
The librational behavior of Epimetheus is reported in Table \ref{tab:libepim}. Contrarily to Janus, the amplitudes of the short-period terms at $\nc \pm\nu$ and $\nc \pm 2\nu$ are greater than the terms at $\nc$. This is due to the fact that $J_1(\beta_1) > J_2(\beta_1)>J_0(\beta_1)$ for Epimetheus, while the relation $J_0(\beta_1)>J_1(\beta_1)>J_2(\beta_1)$ holds for Janus (see Tables \ref{tab:JJanus} and \ref{tab:JEpim}). 
From comparing the first two columns of Table \ref{tab:libepim} we find that the accuracy on the short-period terms obtained for Epimetheus is worse than in the case of Janus.  It reaches about $9\%$ for $ p =1$ and increases to $28\%$ for $p= 4$ and even more  than  for $80\%$ for $p\ge 5$. As for Janus, the accuracy of the analytical solution would be strongly increased if the terms related to $\beta_3$ and possibly  to $\beta_5$ were taken into account. 
\prep{
\tableXlibepim
}

\subsection{Adiabatic-like solution of the equation of the libration in longitude}
\label{sec:adiab}
 We have seen in section \ref{sec:dyn-eq}, neglecting the quadratic terms in eccentricity, that the short-period component of $\gamma$ can be approximated by Eq. (\ref{eq:approx1}), which is the equation 
\be
\ddot y + \sigb^2 y = 2\eb\sigb^2\sin\ell
\label{eq:linn}
\ee
when $N$, in the series $\cS$, tends towards the infinity.
To solve (28) we have expanded $y$ in Fourier series, here we use the  fact  that two time scales coexist in the problem in order to solve the equation (37). 
The frequency $\nu$ being very small with respect to $\nc$, it is possible to expend the solution in Taylor series of the small parameter $\eps = \nu/\nc$.   
First, let us remember that, according to (\ref{eq:ell}), the mean anomaly $\ell$ can be written: $\ell = \nc t + \cB(\nu t)$ were $\cB$ contains the long-time variations associated to the orbital swap\footnote{The subscript $N$ has been removed from $\cB$ to indicate that the summation can be infinite}. 

 If we now introduce  the three angles $(\theta_1,\theta_2,\theta_3)$, defined by $\theta_1 =  \nc t$,  $\theta_2 = \nu t$ and $\theta_3 = \sigb t$, we can see any solutions of (\ref{eq:linn}), as a quasiperiodic function of the three previously defined angle. More precisely, this equation being linear, its solution reads: 
 \be
y(\theta_1,\theta_2,\theta_3) =  \tilde y(\theta_1,\theta_2) + \hat y(\theta_3) 
\ee
 $\hat y$ being periodic and  $\tilde y$ quasiperiodic.
 The forced solution is consequently $\tilde y$. Therefore, in order to obtain this particular solution, it is enough to restrict our analysis to the set of the quasiperiodic functions of the two variables $(\theta_1,\theta_2)$. Consequently, if $y$ is an element of this set, its time derivative reads:
 \be
 \Der y t= \nc \Dron {y}{\theta_1} +  \nu \Dron {y}{\theta_2}
 \label{eq:deriv}
 \ee
 
It turns out that the forced solution of the equation (\ref{eq:linn}) is also solution of the linear partial differential equation:
 \icar
 {
 \be
 \nc^2\frac{\partial^2 y}{\partial \theta_1^2}
 + 2\nc\nu\frac{\partial^2 y}{\partial \theta_1\partial \theta_2}
 + \nu^2 \frac{\partial^2 y}{\partial \theta_2^2}
 +\sigb^2 y = 2\sigb^2\eb\sin(\theta_1 + \cB(\theta_2))
 \label{eq:pde}
 \ee 
 }
\prep
 {
 \be
\begin{split} 
 \nc^2\frac{\partial^2 y}{\partial \theta_1^2}
 + 2\nc\nu\frac{\partial^2 y}{\partial \theta_1\partial \theta_2}
 + &\nu^2 \frac{\partial^2 y}{\partial \theta_2^2}
 +\sigb^2 y =   \\
 &2\sigb^2\eb\sin(\theta_1 + \cB(\theta_2))
 \end{split} 
 \label{eq:pde}
 \ee
 }

 If we now expand the solution $y$  in Taylor series with respect to $\eps$, that is: $y = y_0 + \eps y_1 + \cdots + \eps^p y_p +\cdots $, the functions $y_p$  satisfy the sequence of differential equations: 
\icar{
\begin{align} 
\label{eq:pert0}
y_0: \quad\quad & \nc^2\frac{\partial^2 y_0}{\partial\theta_1^2} + \sigb^2 y_0 = 2\sigb^2\eb\sin(\theta_1 + \cB(\theta_2)) \\
\label{eq:pert1}
y_1: \quad\quad  & \nc^2\frac{\partial^2 y_1}{\partial\theta_1^2} + \sigb^2 y_1 = -2\nc^2\frac{\partial^2 y_0}{\partial\theta_1\partial\theta_2} \\
\label{eq:pertp}
y_p, \, \forall p \ge 2: \quad\quad  
& \nc^2\frac{\partial^2 y_p}{\partial\theta_1^2} + \sigb^2 y_p = -2\nc^2\frac{\partial^2 y_{p-1}}{\partial\theta_1\partial\theta_2} -\nc^2\frac{\partial^2 y_{p-2}}{\partial\theta_2^2} 
\end{align} 
}

\prep{
\begin{align} 
\label{eq:pert0}
y_0: \quad\quad & \nc^2\frac{\partial^2 y_0}{\partial\theta_1^2} + \sigb^2 y_0 = 2\sigb^2\eb\sin(\theta_1 + \cB(\theta_2)) \\
\label{eq:pert1}
y_1: \quad\quad  & \nc^2\frac{\partial^2 y_1}{\partial\theta_1^2} + \sigb^2 y_1 = -2\nc^2\frac{\partial^2 y_0}{\partial\theta_1\partial\theta_2} \\
\label{eq:pertp}
y_p : \quad\quad  
& \nc^2\frac{\partial^2 y_p}{\partial\theta_1^2} + \sigb^2 y_p = -2\nc^2\frac{\partial^2 y_{p-1}}{\partial\theta_1\partial\theta_2} \\
\phantom{y_p : \quad\quad} & \phantom{ \nc^2\frac{\partial^2 y_p}{\partial\theta_1^2} + \sigb^2 y_p =} -\nc^2\frac{\partial^2 y_{p-2}}{\partial\theta_2^2},  \quad \forall p \ge 2 \nonumber
\end{align} 
}

Solving iteratively these equations, it is easy to show that the forced solution of the libation equation reads:
\icar{
\be
y(t) = \frac{2\eb\sigb^2}{\sigb^2 - \nc^2}
   \left[
         \left( 1 + \eps U(\nu t,\eps)\right) \sin\left(\nc t +\cB(\nu t)\right)   +
             \eps^2 V(\nu t,\eps) \cos\left(\nc t +\cB(\nu t)\right)
   \right]
\ee
}
\prep{
\be
\begin{split}
y(t) = \frac{2\eb\sigb^2}{\sigb^2 - \nc^2}  \big[
         ( 1 &+ \eps U(\nu t,\eps) ) \sin (\nc t +\cB(\nu t) )    \\ 
         & \eps^2 V(\nu t,\eps) \cos\left(\nc t +\cB(\nu t)\right)  \big]
\end{split}
\ee
}

where $U$ and $V$ are periodic functions in $\nu t$ that can be expanded in power series of  $\eps$ as:
\be 
U = \sum_{p\ge0} \eps^p U_{p+1} ,\, \text{and}\quad V = \sum_{p\ge0} \eps^p V_{p+2}
\ee
The coefficients $U_p$ and $V_p$, which depend on $\cB$ and on it first $p$ derivatives, are deduced form (\ref{eq:pert0}) to (\ref{eq:pertp}) by induction. 
The approximation of zero order (when $\nu$ is neglected with respect to $\nc$):
\be
y_0(t) = \frac{2\eb\sigb^2}{\sigb^2 - \nc^2}\sin\left(\nc t +\cB(\nu t)\right) = \frac{2\eb\sigb^2}{\sigb^2 - \nc^2}\sin\ell
\label{eq:sol0}
\ee
 is the solution that we intuitively get freezing the long-term temporal variations.  
 This solution is close to the one used in  \citet{TiThBu2009} for the libration, that is: $A\sin(n(t) \, t +\phi)$, where both amplitude $A$ and phase $\phi$ have been fitted  to the observations. In this expression, the instantaneous mean motion $n(t)$ reflects, like $\cB(\nu t)$ in formula (\ref{eq:sol0}),  the long-term variations of the "instantaneous" mean motion  due to the orbital swap.

 Even if the solution $y_0$ given by (\ref{eq:sol0}) resembles the libration in the Keplerian case (section \ref{sec:KO}), their behaviors are very different. Indeed, in the Keplerian case the mean anomaly $\ell$ increases linearly with the time, while in the more realistic case that we are considering, a large 8-years periodic motion in superimposes to this linear evolution (about $36^\circ$ for Janus and more than $130^\circ$ for Epimetheus). Forget this long-term variation could lead to an error on the amplitude of the libration reaching at worst $50\%$ for Janus and $100\%$ for Epimetheus.

 In order to evaluate the accuracy  of the "frozen" solution $y_0$ given by (\ref{eq:sol0}), we have to estimate, at least, the size of the term $U_1$ which appears in the expression of $y_1$. Substituting $y_0$ and solving the differential equation (\ref{eq:pert1}), we have:
\prep{
 \be
 \begin{split}
 y(t) = \frac{2\eb\sigb^2}{\sigb^2 - \nc^2}
& \left(
 1 + 2\eps\frac{\nc^2}{\sigb^2 - \nc^2} \cB'(\nu t) 
 \right)\times \\
 &\sin\left(\nc t +\cB(\nu t)\right)  +O(\eps^2)
  \end{split}
 \ee
}
\icar{
 \be
 y(t) = \frac{2\eb\sigb^2}{\sigb^2 - \nc^2}
 \left(
 1 + 2\eps\frac{\nc^2}{\sigb^2 - \nc^2} \cB'(\nu t) 
 \right)
 \sin\left(\nc t +\cB(\nu t)\right)  +O(\eps^2)
 \ee
}

 where $\cB'$ is the first derivative of $\cB$.

Deducing from numerical simulations that the upper bounds of   $\vert \cB'\vert$ are respectively $1.5$ for Epimetheus and $0.45$ for Janus, it turns out that the addition of $y_1$ in the forced solution, modified the amplitudes of the "frozen" solutions of about $0.64 \%$ for Epimetheus and  $0.03 \%$ for Janus, which is far lower than the accuracy of the observations.

\section{Discussion}
\label{sec:discuss}
\subsection{Higher harmonics}

\citet{TiThBu2009} found in the shape fitting residuals an unexplained offset in the direction of the longest figure axis of the moons,  $5.2^{\circ}$ in the case of Janus. For Epimetheus, the detected offset is within the error bars. These authors suggested that the departure observed for Janus is due to large lateral density anomalies. We investigate such a hypothesis by assuming that the satellites shapes depart from triaxial, hydrostatic shapes, due to mass anomalies expressed at the third degree of spherical harmonics  (as for the Moon e.g. \citealt{Eck1981}). In this case, in addition to periodic terms and small shifts in the proper frequency, a constant term appears in the dynamical Equation leading to a constant offset of the form
\begin{equation}
\gamma_{\rm{hs3}} = \frac{ \zeta}{\sigb^2} (-15 S_{33} + 0.5 S_{31})
\label{hs3}
\end{equation}
where $\zeta = \left(\frac{3n^2}{C}\right) \left(\frac{R}{a}\right)  = 0.308$ rad/days$^2$ for Janus and $0.198$ rad/days$^2$ for Epimetheus, where $S_{33}$ and $S_{31}$ are spherical harmonics of order 3 and degree 3 and 1, respectively. In the expression~(\ref{hs3}), we have assumed that the spherical harmonics $C_{13}$ and $C_{33}$ are negligible with respect to $C_{22}$. Thus, the offset can be expressed as
\begin{equation}
\gamma_{\rm{hs3}}^J = 0.72^{\circ} (-15 S_{33} + 0.5 S_{31})
\end{equation}
for Janus and 
\begin{equation}
\gamma_{\rm{hs3}}^E = 0.16^{\circ} (-15 S_{33} + 0.5 S_{31})
\end{equation}
for Epimetheus.

As a consequence, an offset of $5.2^{\circ}$ requires that the combination $(-15 S_{33} + 0.5 S_{31})$ is about $7$ for Janus, which seems very large even for rubble-piles. In this case, the shape would be far from an ellipsoid.  Nevertheless, the contribution of density anomalies cannot be completely ruled out, and more complex shape models remain to be developed in order to better assess the influence of non-hydrostatic anomalies on rotation.

\subsection{Tidal dissipation}
\label{sec:tidal}

A second possible origin of offset determined by \citet{TiThBu2009} might be related to the tidal torque. Indeed, Saturn raises a tidal bulge on each moon that is shifted from the planet-satellite direction due to the inelastic response of the moon material. Saturn exerts a gravitational torque on this tidal bulge and the body responds by displacing its permanent bulge so that it cancels the average saturnian torque acting on the tidal bulge. Such a displacement has been measured for the Moon (Williams et al. 2001) and estimated for Enceladus (Rambaux et al. 2010). First, we analytically introduced the main tidal deformation as a supplementary term in equation (\ref{eq:linn}), describing the adiabatic behavior of the moons, of the form 
\begin{equation}
T = -k_2 R^5 \frac{3 GM_S^2}{a^6} ( U_{11}U^*_{12} - U_{12}U^*_{11}),
\end{equation}
where $R$ is the satellite's radius, $k_2$ the tidal Love number, and $U_{ij}=\left(\frac{a}{r}\right)^3 u_i u_j$. The $u_i$ are the direction cosines between Saturn and the satellite. The symbol star means the position of Saturn at the constant time delay $\delta t$ resulting from the dissipation. By taking advantage of the spin-orbit synchronous resonance the quantities $u_1$ and $u_1^*$ are of the order of the unity and $u_2$ and $u_2^*$ are small. Then, we develop $u_2^*$ in Taylor series with respect to $\delta t$. Finally, the deformation included in the adiabatic equation Eq. (\ref{eq:pert0}) reads
\be
\ddot y + 2\lambda \dot y + \sigb^2 y = 2\eb\sigb^2 \sin\ell + 4\eb \nc \lambda \cos\ell
\label{eq:gammalindissip}
\ee
with the dissipative rate 
\be
2 \lambda = \frac{3 k_2 R^3}{C} \frac{\nc^4}{Gm} \delta t
\ee
where $m$ is the mass of the satellite. 
 We assume that the delay is constant and equal to: $\delta t = \nc^{-1}Q^{-1}$.  Following the perturbative analysis performed in section \ref{sec:dyn-eq}, 
we solve this equation by written the particular solution in the form 
$$ y = y_s \sin \ell + y_c \cos \ell$$ 
where $y_s$ is the amplitude of the in-phase term and $y_c$ is the amplitude of the out-of-phase term raises by dissipation. The last term induces a small displacement each time that $\ell = 0 \textrm{ mod(}2\pi)$, of
\be 
y_c= \frac{\delta_1\frac{k_2}{Q} }{\left(1 + \delta_2\left(\frac{k_2}{Q}\right)^2\right)}
\label{yc}
\ee
with 
\be
\delta_1 = \frac{6\eb}{C} \left(\frac{\nc^2R^3}{Gm}\right) \left(\frac{\nc^2}{\sigb^2-\nc^2}\right)^2
\ee
and 
\be
\delta_2 = \frac{9}{C^2} \left(\frac{\nc^2R^3}{Gm}\right)^2 \left(\frac{\nc^2}{\sigb^2-\nc^2}\right)^2
\ee
The coefficients $\delta_1$ and $\delta_2$ are equal to $-0.011$ and $0.3205$ radians for Janus and $-0.603$ and $11.863$ radians for Epimetheus. 
$y_c$ depends on the rheology of the satellite through $k_2/Q$. This ratio depends on the internal structure and therefore on the origin of the bodies. Charnoz et al. (2010) have suggested that many small satellites of Saturn, and especially Janus and Epimetheus, come from the accretion of ring material in the form of lumps that separate from the rings. We expect this accretion scenario to yield homogeneous bodies whose composition is mostly water ice. For the two small moons the Love number $k_2$ 
may be computed from the following relationship (e.g., McDonald 1964):
\begin{equation}
k_2=\frac{3}{2}\frac{1}{1+\frac{19\mu_{r}}{2\rho g R}},
\label{k2}
\end{equation}
where $\rho$ is the mean density for the satellite, $R$ its mean radius, $g$ the average surface gravity, and $\mu_{r}$ the effective shear modulus characteristic of the rubble material. We infer from Goldreich and Sari (2009) that the effective shear modulus of Janus is about 0.3 GPa and that of Epimetheus is  0.2 GPa. Such a low effective modulus implies an increased value of $k_{2}$ with respect to that expected for a monolith, of the order of $4.8\times10^{-4}$ and $3.3\times10^{-4}$ for Janus and Epimetheus, respectively.

With regard to the dissipation factor, its value for aggregates is poorly constrained and depends on the dissipative mechanism acting inside the satellites. Then a dissipation factor of the order of 10 to 10$^{3}$ is a possible range of values for Janus and Epimetheus. The smallest dissipation factor ($Q=10$) leading to the largest displacements of the axis of figure of both satellites and implies an $y_c$  about $3.02\times 10^{-5}$ degrees for Janus and  $1.14\times 10^{-3}$ degrees for Epimetheus, which is too small to explain the offset observed by \citet{TiThBu2009}.

\subsection{Influence of the triaxiality on the librational amplitudes}

\prep{
\figureXtriax
}

The short-period librations are of geophysical interest because their amplitudes depend on triaxiality, as shown in Eq.~(\ref{eq:sol_forced}). We range the triaxiality inside the error bars provided by \citet{TiThBu2009} for Janus $0.100 \pm 0.012$ and for Epimetheus $[0.269:0.315]$. For Janus, the libration amplitude depends linearly on triaxiality, while for Epimetheus this dependence is hyperbolic because the proper period and the orbital period are close to each other (see Fig. ~\ref{fig:triax}).

The model suggested by \cite{Charnoz2009} leads to homogeneous bodies. Alternatively, Porco et al. (2007) suggested that both satellites could have accreted rings particles around a core of satellite material with a lower porosity. This idea is supported by the fact that accretion models can account for the very oblate shapes of the satellites. That model implies a contrast in density between the core and a very porous outer layer. Porosity in that layer could be as large as 60 or 70\%, as has been suggested for comets. There is little constraint on the thickness of that layer. Assuming JanusÕ core is made up of solid, pure water ice, it would have a mean radius of about 70 km. Assuming an end-member model with a solid core of water ice and a 60\% porous outer layer, decreases the mean moment of inertia by about 15\% with respect to the value for a homogeneous body. Unfortunately, the size of the error bars on the triaxiality from \citet{TiThBu2009} is of the order of 15\%, which prevents further investigation of a possible relationship between libration amplitude and internal structure.

\section{Conclusion}

In this paper we have investigated the librational motion of the co-orbital satellites Janus and Epimetheus by using three methods: (1) a perturbative technique based on quasi-periodic expansions, (2) an adiabatic invariant approach by expanding in power series of the small parameter $\nu/\nb$, and (3) a numerical integration. With the perturbative technique, we have detailed the librational behavior. For both satellites the solutions are composed of long-period librations linked to the orbital swap and short-period librations related to the orbital period. 
We found that the amplitudes of the short-period librations depend on the magnitude of the forcing and the proximity to the resonance, as for the librations analyzed in a Keplerian framework, but also on Bessel functions of the amplitude of orbital libration of the moons mean longitudes along their horseshoe orbit. These amplitudes bear the signature of the mass distributions in the satellites and are crucial to investigate the internal structure signature of these objects. On the other hand, the amplitudes related to the long-period librations do not contain any information on the distribution of mass. 
The numerical integration allows us to assess the accuracy of the perturbative development. The accuracy of the analytical solution is good for Janus but poor for Epimetheus because its mass is smaller than Janus and therefore its dynamics is more perturbed. In addition, in order to obtain a compact analytical solution easy to manipulate, we have developed an adiabatic approach yielding directly the amplitude of the short periods.

The analytical approaches have been developed in the most general formalism and may be applied for co-orbitals like Telesto, Calypso, Helene and Polydeuces. The adiabatic approach seems a convenient approach to fit the short-period librations to the observations. 



\subsection*{Acknowledgments}
The authors thank M. Tiscareno, one of the reviewers of the present paper, for his numerous comments and useful suggestions. We would also thank J. Laskar and S. Boatto for fruitful discussions.
Part of this work has been conducted at the Jet Propulsion Laboratory, California Institute of Technology, under a contract with the National Aeronautics and Space Administration. All rights reserved. Government sponsorship acknowledged.

\newcommand{\noopsort}[1]{}

\appendix

\section{Expansion in Bessel functions}\label{a:bessel}

The Bessel functions can be defined as Fourier's coefficients of the $2\pi$-periodic function $u \longmapsto e^{i x \sin{u}}$ where $x$ is a real parameter, that is: 
$$
e^{i x \sin{u}} = \sum_{k=-\infty}^{+\infty} J_k(x) e^{i k u}
$$
\icar{
$$ 
\text{with }\quad J_k(x) = \frac{1}{2\pi} \int_0^{2\pi} \exp i (x \sin u - ku) du
 =  \frac{1}{2\pi} \int_0^{2\pi} \cos(x \sin u - ku) du
  $$
}
\prep{
\be
\begin{split}
\text{with }\quad J_k(x) &= \frac{1}{2\pi} \int_0^{2\pi} \exp i (x \sin u - ku) du \\
                                         &=  \frac{1}{2\pi} \int_0^{2\pi} \cos(x \sin u - ku) du
 \end{split}
 \nonumber
\ee
}
  These functions satisfy the two following relations that we use in Section \ref{sec:dyn-eq} : 
$$ 
\text{ for all}\, p \in\NN, \quad J_{-p}(x) = (-1)^p J_p(x)
 $$ 
 $$ 
\text{ for all}\, p \in\NN, \quad J_{p}(x) =  \frac{x^p}{2^pp!}(1 + O(x^2)) 
 $$ 
 
 In addition, we show in Tables \ref{tab:JJanus} and \ref{tab:JEpim} the values of the main  coefficients $J_q(\beta_p)$ that are  greater than $10^{-6}$. These tables are useful to evaluate the accuracy of our analytical solution (see Section \ref{sec:compar}), and also to identify terms capable of increasing the accuracy of the solution.  

\prep{
\tableXJJanus
}

\prep{
\tableXJEpim
}
\newpage

\section{Table of notations}

\prep{
\tableXdefvar
}

\prep{ 
\end{document}
}
\clearpage
 \icar{
\tableXfrefond
}
\clearpage

\icar{
\tableXQP_relat
}
\clearpage

\icar{
\tableXbulk
}

\clearpage

\icar{
\tableXlibjanus
}
\clearpage

\icar{
\tableXlibepim
}
\clearpage

\icar{
\tableXJJanus
}

\clearpage

\icar{
\tableXJEpim
}

\clearpage

\icar{
\tableXdefvar
}

\clearpage
\prep{
\figureXa
}
\clearpage

\icar{
\figureXa
}
\clearpage

\icar{
\figureXrelat
}

\clearpage

\icar{
\figureXecc
}

\clearpage

\icar{
\figureXerr
}

\clearpage

\icar{
\figureXerrE
}

\clearpage

\icar{
\figureXtriax
}

\icar{
\end{document}
}